\documentclass{pasj00}

\begin{document}
\SetRunningHead{Goto et al.}{Galaxy Clusters at $0.9<z<1.7$ in the AKARI NEP deep field.}
\Received{2008/06/02}
\Accepted{2008/09/28}

\title{Galaxy Clusters at 0.9$<z<$1.7 in the AKARI NEP deep field.}


\author{
Tomotsugu \textsc{Goto},\altaffilmark{1,2}  \thanks{JSPS SPD Fellow}
   Hitoshi \textsc{Hanami},\altaffilmark{3}    
   Myungshin \textsc{Im},\altaffilmark{4}  
Koji  \textsc{Imai},\altaffilmark{5}   
Hanae  \textsc{Inami},\altaffilmark{6}   \\
Tsuyoshi  \textsc{Ishigaki},\altaffilmark{7}   
   Hyung Mok \textsc{Lee},\altaffilmark{4} 
Hideo \textsc{Matsuhara},\altaffilmark{6} 
   Takao \textsc{Nakagawa}\altaffilmark{6}   \\
 Youichi \textsc{Ohyama},\altaffilmark{8} 
   Shinki \textsc{Oyabu},\altaffilmark{6}  
Chris P. \textsc{Pearson},\altaffilmark{9} 
Toshinobu \textsc{Takagi},\altaffilmark{6} \\
and Takehiko \textsc{Wada},\altaffilmark{6} 
} %
 \altaffiltext{1}{Institute for Astronomy, University of Hawaii
2680 Woodlawn Drive, Honolulu, HI, 96822, USA}
\email{tomo@ifa.hawaii.edu}
 \altaffiltext{2}{National Astronomical Observatory of Japan, Mitaka, Tokyo 181-8588}
  \altaffiltext{3}{Physics Section, Faculty of Humanities and Social Sciences, \\ Iwate University, Morioka, 020-8550}  
 \altaffiltext{4}{Department of Physics \& Astronomy, FPRD, Seoul National University, \\Shillim-Dong, Kwanak-Gu, Seoul 151-742, Korea}
 \altaffiltext{5}{TOME R\&D Inc. Kawasaki, Kanagawa 213 0012, Japan}
 \altaffiltext{6}{Institute of Space and Astronautical Science, Japan Aerospace Exploration Agency, \\
   Sagamihara, Kanagawa 229-8510}  
 \altaffiltext{7}{Asahikawa National College of Technology, 2-1-6 2-jo Shunkohdai, Asahikawa-shi, Hokkaido 071-8142 }
 \altaffiltext{8}{Institute of Astronomy and Astrophysics, Academia Sinica, P.O. Box 23-141, Taipei 106, Taiwan, Republic of China}
 \altaffiltext{9}{ISO Data Centre, ESA, Villafranca del Castillo, Madrid, Spain.}  


\maketitle

\begin{abstract}
There is a huge gap between properties of red-sequence selected massive galaxy clusters at $z<1$ and Lyman-break selected proto-clusters at $z>3$. It is important to understand when and how the $z>3$ proto-clusters evolve into passive clusters at $z<1$.

We aim to fill this cluster desert by using the space-based $N4(4\mu m)$ imaging with the AKARI. The $z'-N4$ color is a powerful separator of cluster galaxies at $z>1$, taking advantage of the 4000\AA~break and the 1.6$\mu$m bump. We carefully selected 16 promising cluster candidates at $0.9<z<1.7$, which all show obvious over-density of galaxies and a prominent red-sequence.
 
 At this redshift range, the mid-infrared $S_{15\mu m}/S_{9\mu m}$ flux ratio is an extinction-free indicator of galaxy star formation activity due to the redshifted PAH emission lines (6.2,7.7 and 8.6$\mu$m). 
We show statistically that the cluster galaxies have a lower $S_{15\mu m}/S_{9\mu m}$ flux ratio than field galaxies, i.e., cluster galaxies already have lower star-formation activity at $0.9<z<1.7$, pushing the formation epoch of these galaxy clusters to a higher redshift.
\end{abstract}

\section{Introduction}


 High redshift clusters are highly important objects in variety of science:
 For example, it has been known that galaxy cluster environment strongly affects the evolution of galaxies. The morphology-density relation suggests the cluster environments transform galaxies into early-type \citep{2003MNRAS.346..601G}. The star formation rate of cluster galaxies is significantly lower than in the field \citep{2004AJ....128.2677T}. 
 A galaxy cluster is the most massive virialized system in the Universe; cluster studies provide an important tool for gauging the growth of structure and probing the density of the underlying dark matter and energy 
\citep{2005Natur.435..629S}.  A presence of (even a few) massive clusters at high redshift can constrain cosmological parameters $\Omega_M$ and $\sigma_8$ \citep{1997ApJ...485L..53B}. Cluster studies can illuminate how dark matter haloes collapse and large scale structure forms. High redshift clusters can help constrain feedback processes caused by star-formation and AGN \citep{1998A&A...331L...1S}.

 However, there has been a missing link at $1<z<2$ in the current cluster surveys.
 Traditionally, in the nearby Universe, galaxy clusters are known to form a strong color-magnitude relation (red sequence), dominated by red, early-type galaxies. And thus, galaxy clusters are primarily found by searching for red sequence galaxies clustering on the sky \citep{2000AJ....120.2148G,2002AJ....123.1807G}. This method utilizes the 4000\AA~ break, and thus, optical surveys can only search for clusters at z$<$1 (the limit set by $i-z$ color). This is why only few galaxy clusters are known at $z>1$ today, most of them stemming from X-ray surveys \citep{2005ApJ...623L..85M,2005ApJ...634L.129S,2006ApJ...646L..13S}.
 
 At z$>$3, Ly$\alpha$ emission redshifts into the optical band. People found clustering of Lyman-$\alpha$ emitters (LAE) or Lyman Break Galaxies (LBG), often around a strong radio galaxy or AGN (thus, these are biased surveys). However, these proto-clusters are groups of star-forming galaxies (LAE,LBG) and thus, fundamentally different from the low redshift clusters dominated by red early-type galaxies \citep{2004Natur.427...47M}. It is important to investigate how and when these proto-clusters evolve into passive galaxy clusters in the nearby Universe. 

 Also, due to these selection methods, there inevitably exists strong selection bias between the proto- and classic (passive) clusters. It is important to study the evolution of galaxy clusters using a sample selected in a consistent way. This paper provides one of the attempts to extend the classical cluster search to higher redshift. 

In summary, due to the red/blue wavelength limit of the optical surveys, there has been a hole in galaxy cluster surveys at 1$<z<$3. 
 We aim to fill this missing link in previous galaxy cluster surveys by utilizing the deep near infrared imaging with the $AKARI$ satellite combined with the ground based deep $z'$-band image with the SuprimeCam on board the Subaru telescope. 
 Deep $z'$-band images combined with the $AKARI$ 4$\mu$m images straddle the 4000\AA~break of galaxies at $1<z<2$. And therefore, we can naturally extend the classical 4000\AA-break search of galaxy clusters to this redshift range of the cluster desert.

 Unless otherwise stated, we adopt the WMAP cosmology: $(h,\Omega_m,\Omega_L) = (0.7,0.3,0.7)$ \citep{2008arXiv0803.0547K}. 

\section{Data}

 The AKARI, the Japanese infrared satellite \citep{2007PASJ...59S.369M}, performed a deep imaging in the North Ecliptic Region (NEP) in 2-24$\mu m$ (14 pointings in each field over 0.4 deg$^2$; \cite{2006PASJ...58..673M,2007PASJ...59S.543M,WadaNEP2008}). Due to the solar synchronous orbit of the AKARI, the NEP is the only field with deep imaging in these wavelengths (but see \cite{2005ApJ...634..128T}). Since the Spitzer space telescope does not have sensitivity at $9<\lambda<20\mu$m  (a gap between the IRAC and MIPS) except for the IRS peak up array at 16 $\mu$m and 22 $\mu$m, the AKARI NEP survey will be the only field with deep images in these wavelengths at least for a next decade.
 The depth provided by the AKARI satellite in $N4$-band (4$\mu$m) reaches 8.0 $\mu$Jy in 5 $\sigma$\citep{WadaNEP2008}, which can in principle detect bright galaxies up to $z\sim$1.7. The 5 $\sigma$ sensitivity in other wavelengths ($N2,N3,N4,S7,S9W,S11,L15,L18W$ and $L24$) are 14.2, 11.0, 8.0, 48, 58, 71, 117, 121 and 275$\mu$Jy \citep{WadaNEP2008}.

 A part of the NEP deep field was observed in $z'$-band with the Subaru telescope \citep{2007AJ....133.2418I,WadaNEP2008}, reaching the limiting magnitude of $z_{AB}=$26 mag in one field of view of the Suprime-Cam.
 The combination of deep $z$- and $N4(4\mu m)$-images provides us with a unique opportunity to search for galaxy clusters in the previous cluster desert at $0.9<z<1.7$. The $N4(4\mu m)$-band covers the same area observed with the $N3,N4,S7,S9W,S11,L15,L18W$ and $L24$ bands, allowing us to investigate the mid-infrared properties of discovered clusters immediately.  
 We use the AB magnitude system in this paper.

\section{Method}\label{sec:method}
\subsection{Cluster Detection}\label{sec:detection}

 We have previously developed an efficient galaxy cluster finding algorithm called the Cut \& Enhance method \citep{2002AJ....123.1807G}, which uses color-cuts and spatial clustering information to enhance a signal from galaxy clusters. In this paper we apply this method to $z'-N4$ colors of galaxies in the NEP deep field. 

  In the right panels of Fig.\ref{fig:images}, we show the $z'$ vs. $z'-N4$ color-magnitude diagram. 
 Orange dots are k-corrected, evolutionary models of a passive galaxies with $10^{11}M_{\odot}$ (total of the gas and stellar mass) as a function of redshift (indicated in the figure). This model assumes passive evolution of galaxies.
   5 $\sigma$ detection limit of the $N4$ band is 8.0 $\mu$Jy or 21.67 ABmag. 
  This detection limit of the $N4$ band is the deepest among the AKARI $N2,N3$ and $N4$ bands. In addition, because of the redshifted 1.6$\mu$m bump \citep{2002AJ....124.3050S},  the $N4$ band does not become fainter as quickly as $N2$.  
  As can be easily seen, colors of galaxies become redder and redder in $z'-N4$ color from z=0.9 to z=1.7. 
  As a reference, the overplotted dotted straight lines show the artificially redshifted color-magnitude relation observed in the Coma cluster at z=0.9, 1.0, 1.1, 1.25, 1.5 and 1.7. 
Because at $z>0.9$, $z'$- and $N4$-bands straddle the 4000\AA~ break, the  $z'-N4$ color becomes redder and redder with increasing redshift. In addition,
 the N4 magnitude can take advantage of the 1.6$\mu$m bump and  does not became as fainter as N2 or N3 bands at our targeted redshift range \citep{2008arXiv0804.4798E}; 
 in Fig.\ref{fig:filters}, we show three spectra of instantaneous burst model galaxies redshifted to z=1.3, with ages of 0.5, 1, and 10 gigayears of age with Salpeter initial mass function and the solar metallicity \citep{2003MNRAS.344.1000B}. The overlaid are the filter response of $z'$, $N2,N3$ and $N4$ bands. At z=1.3, the redshifted 1.6$\mu$m bump is in the $N4$-band.
 This is why in Fig.\ref{fig:images}, the model track becomes redder, but not very fainter with increasing redshift (and thus moves straight up in the figure). 
 Taking advantage of this color-shift, we can efficiently find galaxy clusters by using color-cut corresponding to each redshift, and by searching an overdensity of galaxies within a color-cut (i.e., overdensity of galaxies with the same color).

 To find galaxy clusters in this field, we separate galaxies in the $z'-N4$ color space in the bins of $\Delta (z'-N4)$=0.2. The width of this color-cut is set to take account of relatively large errors of the $N4$ magnitude.
 We move this color-cut from blue ($z'-N4$=1.4 targeting $z=0.9$) to red ($z'-N4$=3.7 targeting $z=1.7$) by 0.1 mag at a time as if scanning along the redshift. The adjacent bins have overlap of 0.1 mag in color, assuring that a cluster which happens to be on a border of a color cut is not missed.
 This color-cut removes most of fore- and background galaxies nicely, enhancing the density contrast of galaxy clusters.

 In each color-cut, we compute a local galaxy density of individual galaxies 
by computing a number of galaxies within 1.5 Mpc of radius within the same color-cut.
 If the galaxy has more than 2$\sigma$ overdensity than the average galaxy density in the color-cut, the galaxy is selected to be in a dense region. 
 We then connect these galaxies using a friend-of-friend algorithm with the linking length of (projected) 0.5 Mpc \citep{2005MNRAS.356L...6G}. If more than 4 galaxies are connected, we select them as a galaxy cluster candidate.
 Sometimes, a same cluster can be found in two adjacent color-cuts. In such cases, we combine such clusters if they were within 1.5 Mpc from each other and in the adjacent color-cuts.

 Using this method, we have found 16 galaxy cluster candidates in the range of $0.9<z<1.7$.
 When tuned for lower redshift range, a known cluster, RXJ175719.4+6633 (z=0.6909; \cite{2003ApJS..149...29G,2006A&A...446...97B,2006ApJS..162..304H}) was successfully detected with our cluster finder, showing our methodology works well.
 The list of these cluster candidates are presented in Table \ref{tab:LTsample}.


%

\subsection{Redshift Estimate}

The next important step is to measure redshift of each cluster we found.  Spectroscopic redshift measurement is ideal. 
However, at  $0.9<z<1.7$, taking spectra of cluster galaxies which usually do not show emission lines is costly. 
Especially, for clusters at $z>1.5$, near-infrared spectroscopy is required to measure Ca H\&K absorption lines. For galaxies close to $N4$=22mag, such near-infrared spectroscopy is impossible with a reasonable amount of existing 8m class telescope time.

 However, cluster galaxies are known to have a tight color-magnitude relation \citep{1998A&A...334...99K}, i.e., galaxy clusters have many passive galaxies, forming a line in the color-magnitude diagram (called the red-sequence). Such red-sequence galaxies have a strong 4000\AA~ break, and thus, their photometric redshift can be measured much more reliably. Since galaxy clusters have many red-sequence galaxies,  it is easier to measure photometric redshift of galaxy cluster than those of individual galaxies. 
 In the past, such cluster photometric redshift measurement reported accuracy of $\delta z <0.02$ using only 2 colors that straddle the 4000\AA break at the redshift (\cite{2000AJ....120..540G,2002AJ....123.1807G,2002ApJ...571..136I}).

We apply this method to our clusters using the $z'-N4$ color. 
First, we fit a line with the slope of the Coma color-magnitude relation to cluster member galaxies within the $z'-N4$ color-cut selected in Section \ref{sec:detection}. This model assumes the passive evolution of galaxies. 
 Then, we re-select member galaxies within $\pm0.1$mag in $z'-N4$ along the fitted line, and within 1.5 Mpc from the cluster center, which are median coordinates of galaxies in the initial selection.
 We re-fit the line again to the new set of member galaxies to determine the photmetric redshift of each cluster. In most cases, a difference between initial redshift estimate and our final photometric redshift is less than 1\%.

\section{Results}
\label{sec:results}


\subsection{RXJ175719.4+663131}

In the AKARI NEP field, there is a known galaxy cluster, RXJ175719.4+663131, at z=0.6909.
The ROSAT X-ray satellite has a similar sun synchronous orbit as the AKARI, and thus has deepest data around the NEP region. This cluster was first found by the ROSAT satellite in X-ray \citep{2003ApJS..149...29G,2006ApJS..162..304H,2006A&A...446...97B}. The redshift of this cluster based on 5 spectroscopic member galaxies is z=0.6909.
  
 In the upper panel of Fig.\ref{fig:RXJ1757_image}, we show $z'$- (left) and $N4$-(right) band images of this cluster. 
Galaxies marked with red circles have $z'-N4$ color consistent to be at redshift of 0.6909. It is seen that these galaxies are clustered within the 1.5 Mpc around the cluster center. In addition to galaxies marked with red circles, there are obvious overdensity of galaxies in the $z'$-band image, showing this is truly a galaxy cluster.

 In Fig.\ref{fig:RXJ1757_cmd}, we show $z'-N4$ vs $z$ color-magnitude diagram of cluster galaxies with correct $z'-N4$ color within the 0.5 Mpc with large dots. 
 A tight color-magnitude relation can be seen.
 The gray dots show all the galaxies within 0.5 Mpc from the cluster center \citep{2003ApJS..149...29G}.   
 This cluster has also been successfully detected with our blind search for galaxy clusters.

\subsection{Most promising cluster candidates}\label{sec:15clusters}


In this section, we present detected cluster candidates at $0.9<z<1.7$.

 We have detected 22 cluster candidates in this redshift range in total using the method described in Section \ref{sec:method}.  Among them, four clusters are identical to the detection in the adjacent color-cut, and thus are merged into single clusters. 
   Among the remaining 18 clusters, one was affected by an artificial spike of the $N4$ image. Another at z=0.97 conincides with the position of the z=1.36 cluster, and can be considered as bluer galaxies of the background cluster.  Therefore, the two candidates are removed from the list.
 In the left panels of Fig. \ref{fig:images}, we show Subaru $z'$-band images of all 16 cluster candidates. 
 The image scale is indicated in the green bar in the figure.
 Galaxies in red circles have a $z'-N4$ color consistent with the red sequence at each redshift. Around these galaxies, there are obvious overdensity of galaxies including galaxies not detected with the AKARI, and thus, these candidates are likely to be real galaxy clusters. Some of cluster candidates even have a possible cD galaxy.  
 In the central panels of Fig. \ref{fig:images}, we show corresponding AKARI $N4$ images of clusters.
 We show a corresponding color-magnitude relation of these cluster candidates in the right panels of Fig.\ref{fig:images}. 
 Filled circles are galaxies in $\delta (z'-N4)<0.2$ from the best-fit color of the red-sequence, and within 1.5 Mpc from the cluster center.
 The gray dots are all galaxies within 1.5 Mpc around the cluster in the NEP deep SuprimeCam field. 
 The orange dots are k-corrected, evolutionary models of a passive galaxies with $M=10^{11}M_{\odot}$ (total of the gas and stellar mass) as a function of redshift (indicated in the figure). The model shows that $N4$ magnitude does not become much fainter in this redshift range because of the 1.6$\mu$m bump.
 

%

 To quantify the significance of the overdensity of cluster candidates in Table \ref{tab:LTsample}, 
we have measured a local overdensity, $\delta_{\Sigma}=\Sigma_{cluster}/\overline{\Sigma_{field}}$,  of cluster cores (galaxies within 0.5 Mpc of radius around the median position of cluster galaxies). 
 Here $\Sigma_{cluster}$  is a number density of cluster galaxies within 0.5 Mpc of radius and in the color-cut.  $\Sigma_{field}$ is computed for each galaxies outside of the cluster in the same way as cluster galaxies (around 0.5 Mpc and in the color-cut), then we take mean to compute $\overline{\Sigma_{field}}$.
The mean field galaxy density was computed using all galaxies except those belong to the cluster in the color-cut corresponding to the cluster redshift. The results are presented in Table \ref{tab:LTsample}. The mean $\delta_{\Sigma}$ is 5.5, which means the galaxy density of cluster cores is 5.5 times larger than that of the corresponding field galaxies.

   We also measure the significance of overdensity,  $\sigma_{\Sigma}=(\Sigma_{cluster}-\overline{\Sigma_{field}})/\sigma(\Sigma_{field})$. We measure  fluctuation of  $\Sigma_{field}$ in $\sigma(\Sigma_{field})$. The results are presented in  Table \ref{tab:LTsample}.  The mean  $\sigma_{\Sigma}$ is 2.4, which statistically occurs with less than 2\% of probability.
 We do not claim our sample is completely free from the contamination. True identity of the clusters needs to be verified spectroscopically.


\begin{figure*}
  \begin{center}
\FigureFile(90mm,80mm){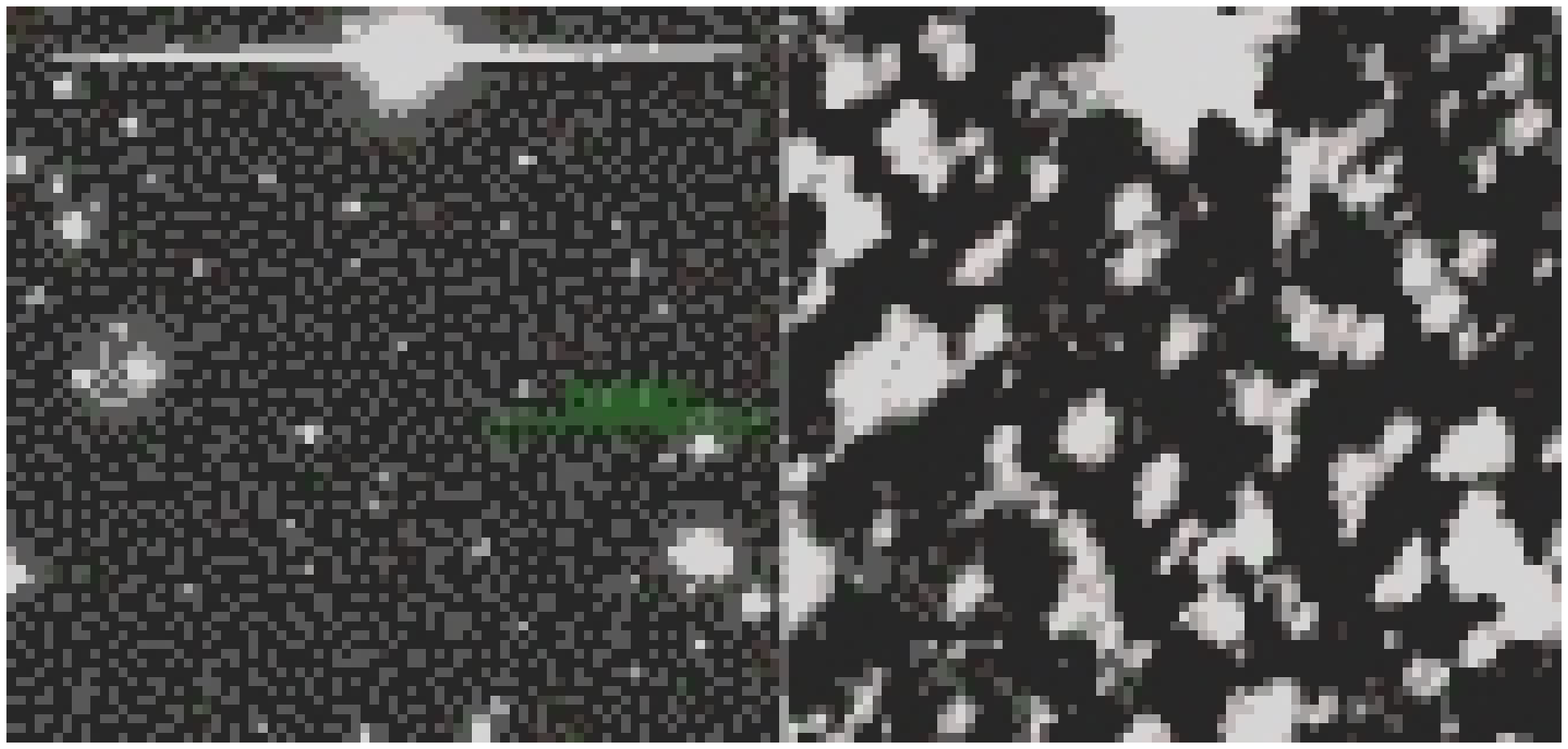}
\FigureFile(75mm,80mm){080822_nep_photoz.ps_pages14}
\FigureFile(90mm,80mm){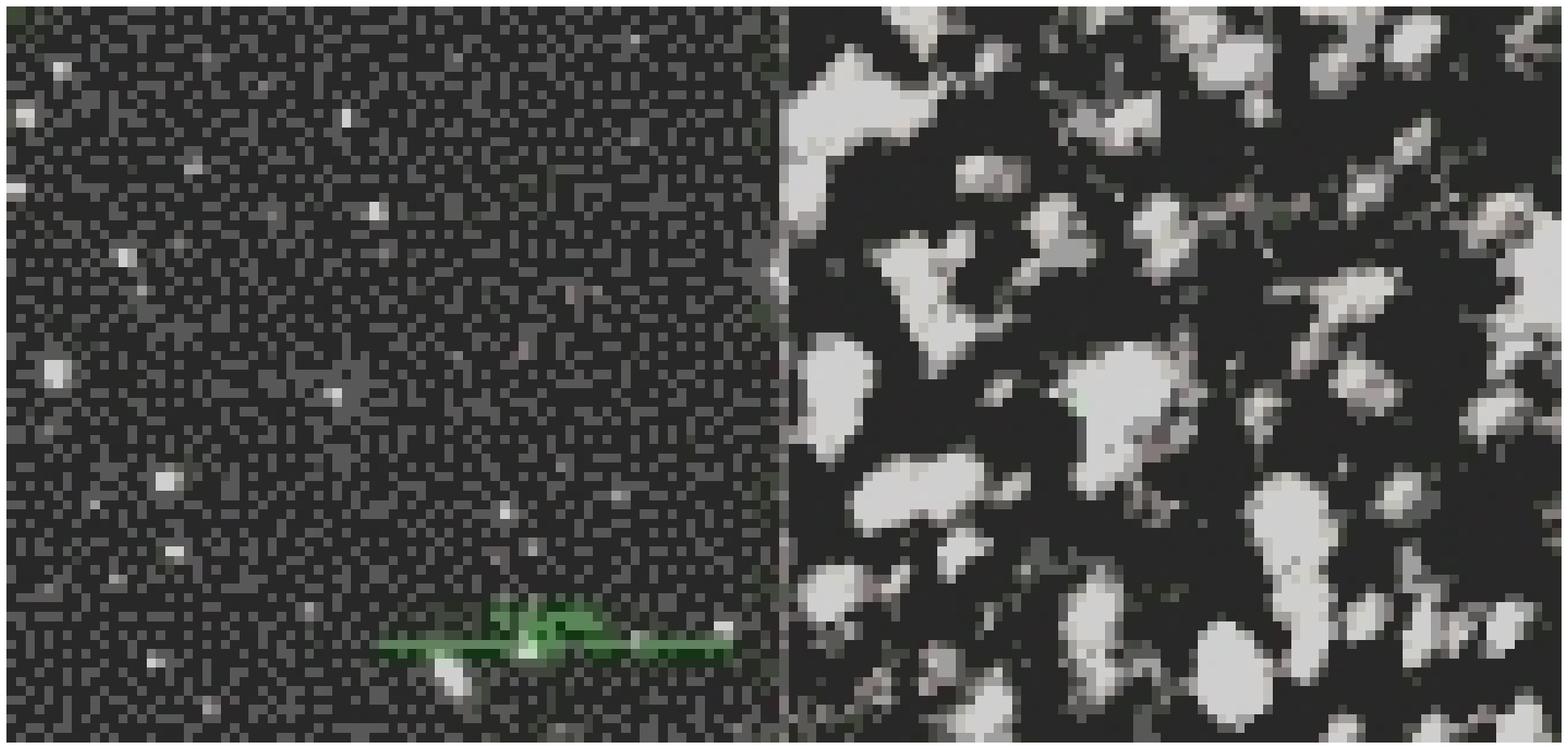}
\FigureFile(75mm,80mm){080822_nep_photoz.ps_pages15}
\FigureFile(90mm,80mm){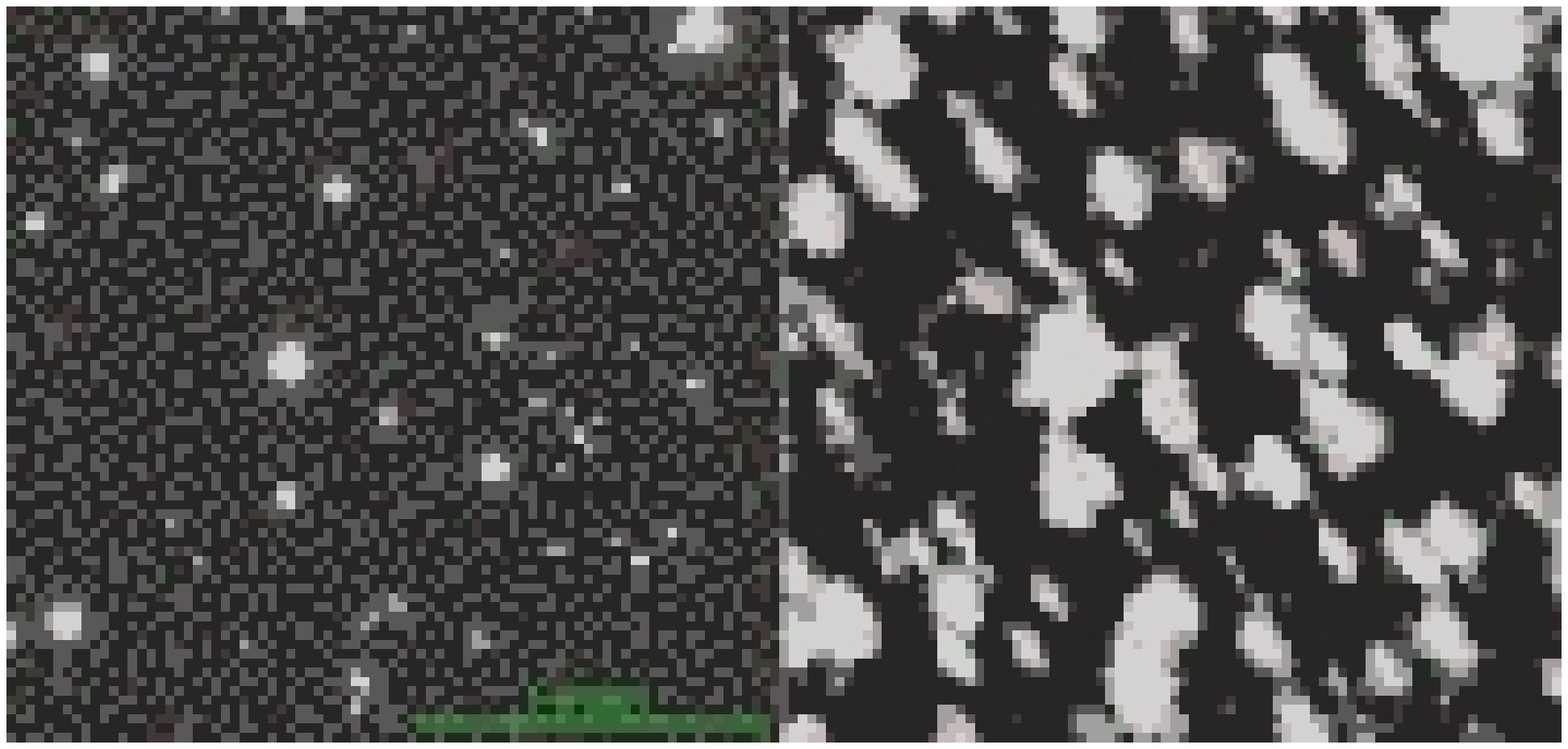}
\FigureFile(75mm,80mm){080822_nep_photoz.ps_pages16}
\FigureFile(90mm,80mm){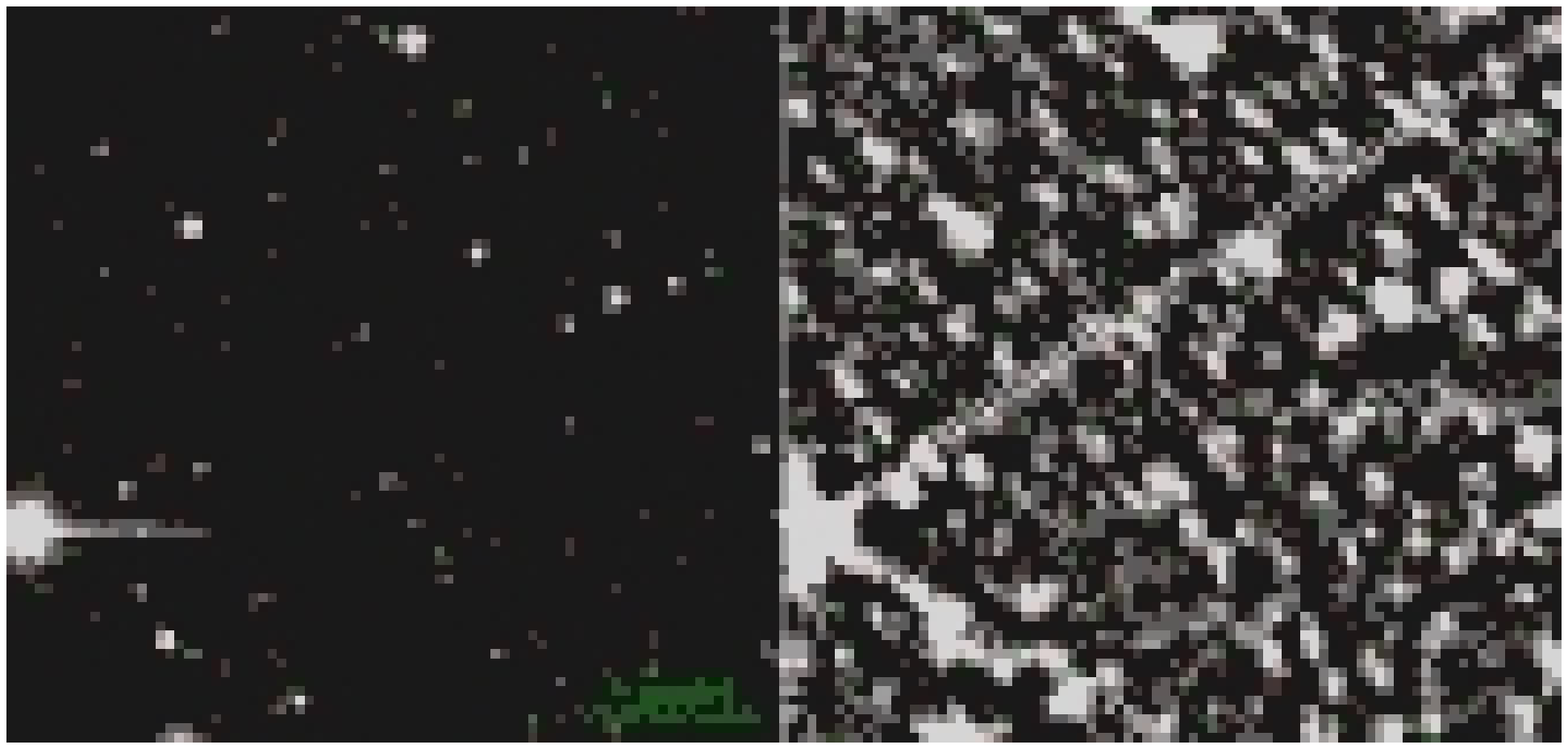}
\FigureFile(75mm,80mm){080822_nep_photoz.ps_pages17}
\end{center}
\caption{ Cluster candidates at $0.9<z<1.7$ found in the NEP deep field. The left panels are Subaru $z'$-band images. The middle panels are AKARI $N4$-band images. The scale is indicated in the left panels. Galaxies marked with red circles (2'' radius) have consistent $z'-N4$ color to be on the red-sequence at the detection redshift. 
 The right panels are $N4$ vs $z'-N4$ color-magnitude diagram, where large black dots are cluster member galaxies. The orange dots show model galaxy colors with evolution as a function of redshift. The red solid line shows the color-magnitude relation of the Coma cluster at the detection redshift.
  The green triangles are galaxies which have photometric redshifts consistent to be at a cluster redshift.  
 For reference, the small gray dots in the background are all galaxies within 1.5 Mpc from the cluster centre.
}\label{fig:images}
\end{figure*}
\setcounter{figure}{0}
\begin{figure*}
\begin{center} 
\FigureFile(90mm,80mm){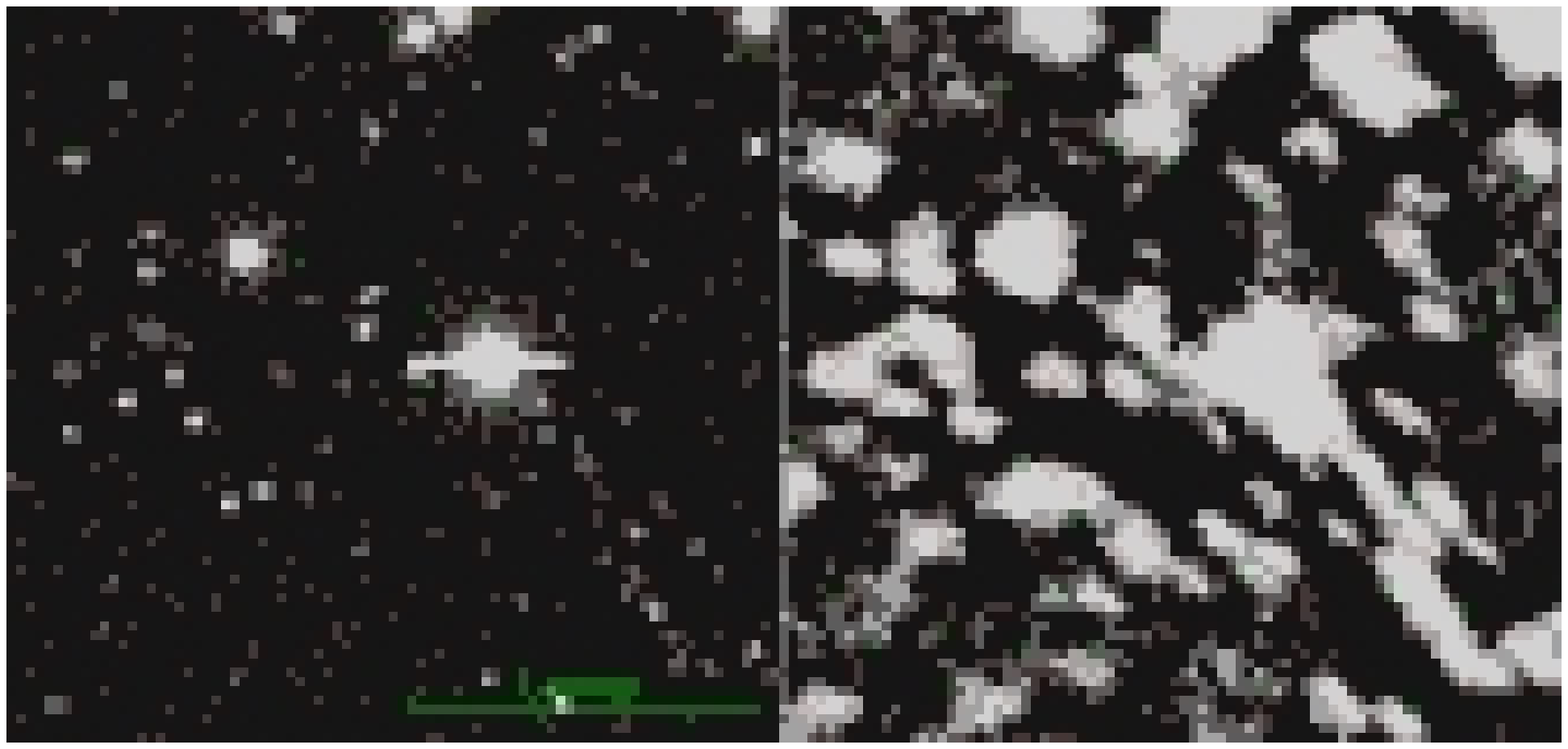}
\FigureFile(75mm,80mm){080822_nep_photoz.ps_pages18}
\FigureFile(90mm,80mm){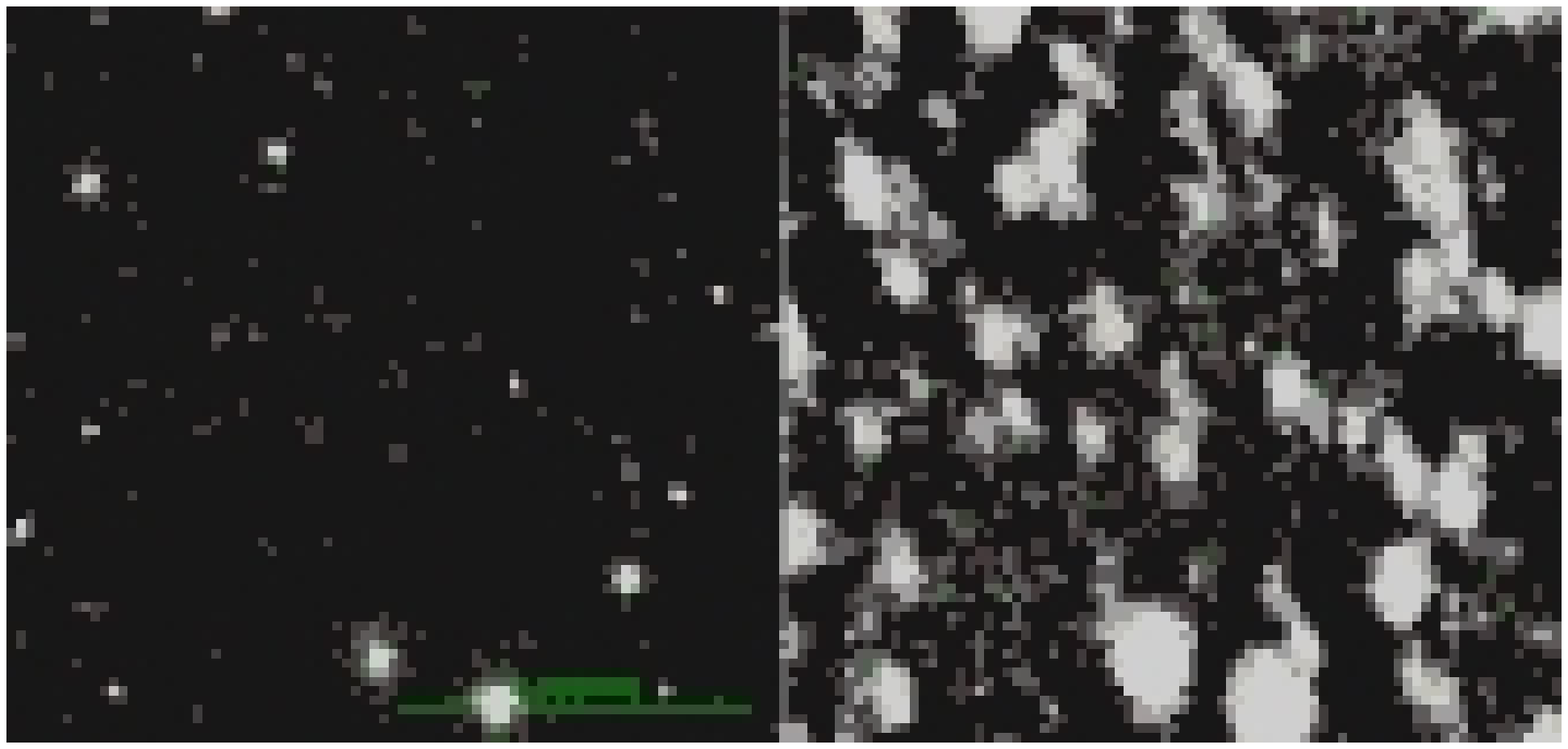}
\FigureFile(75mm,80mm){080822_nep_photoz.ps_pages19}
\FigureFile(90mm,80mm){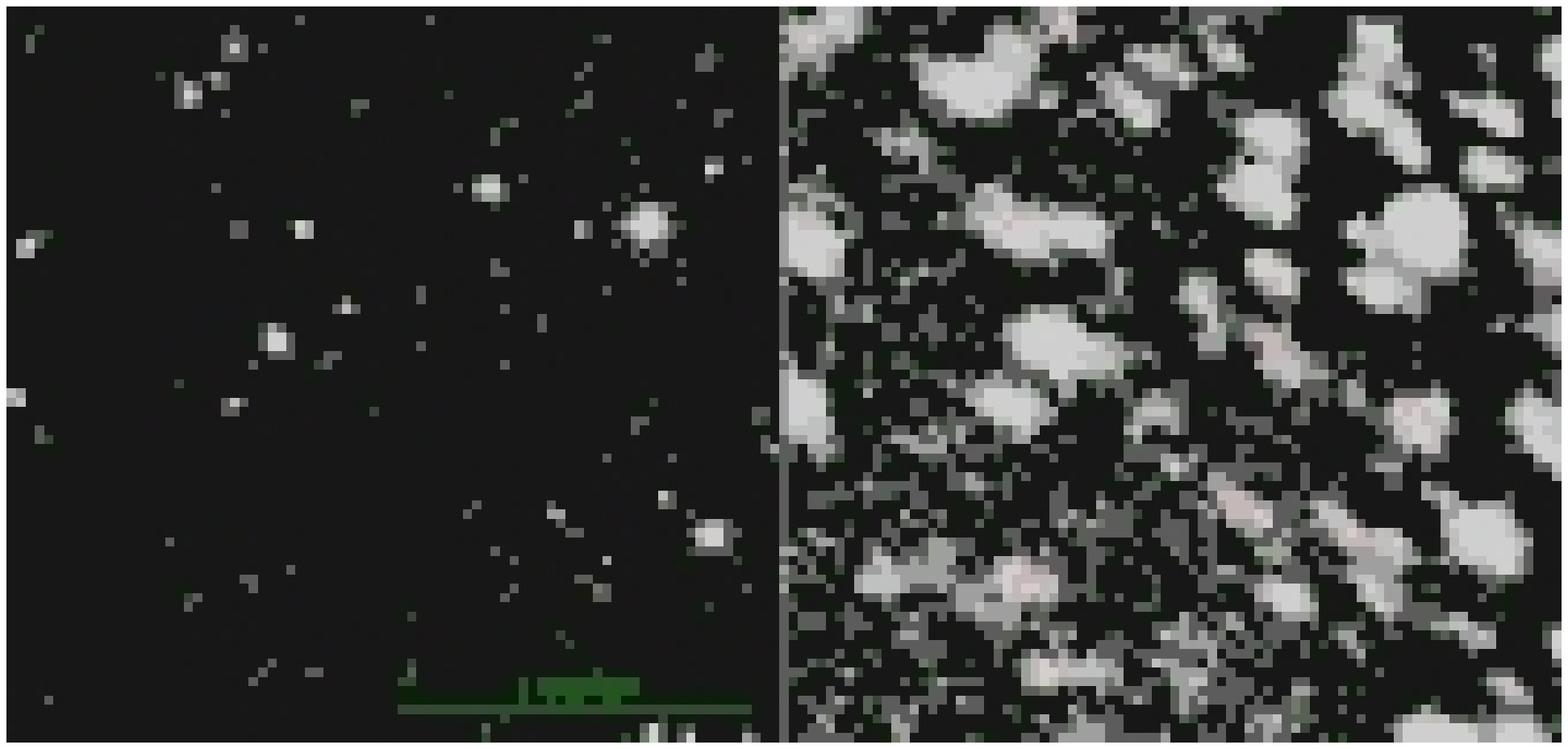}
\FigureFile(75mm,80mm){080822_nep_photoz.ps_pages20}
\FigureFile(90mm,80mm){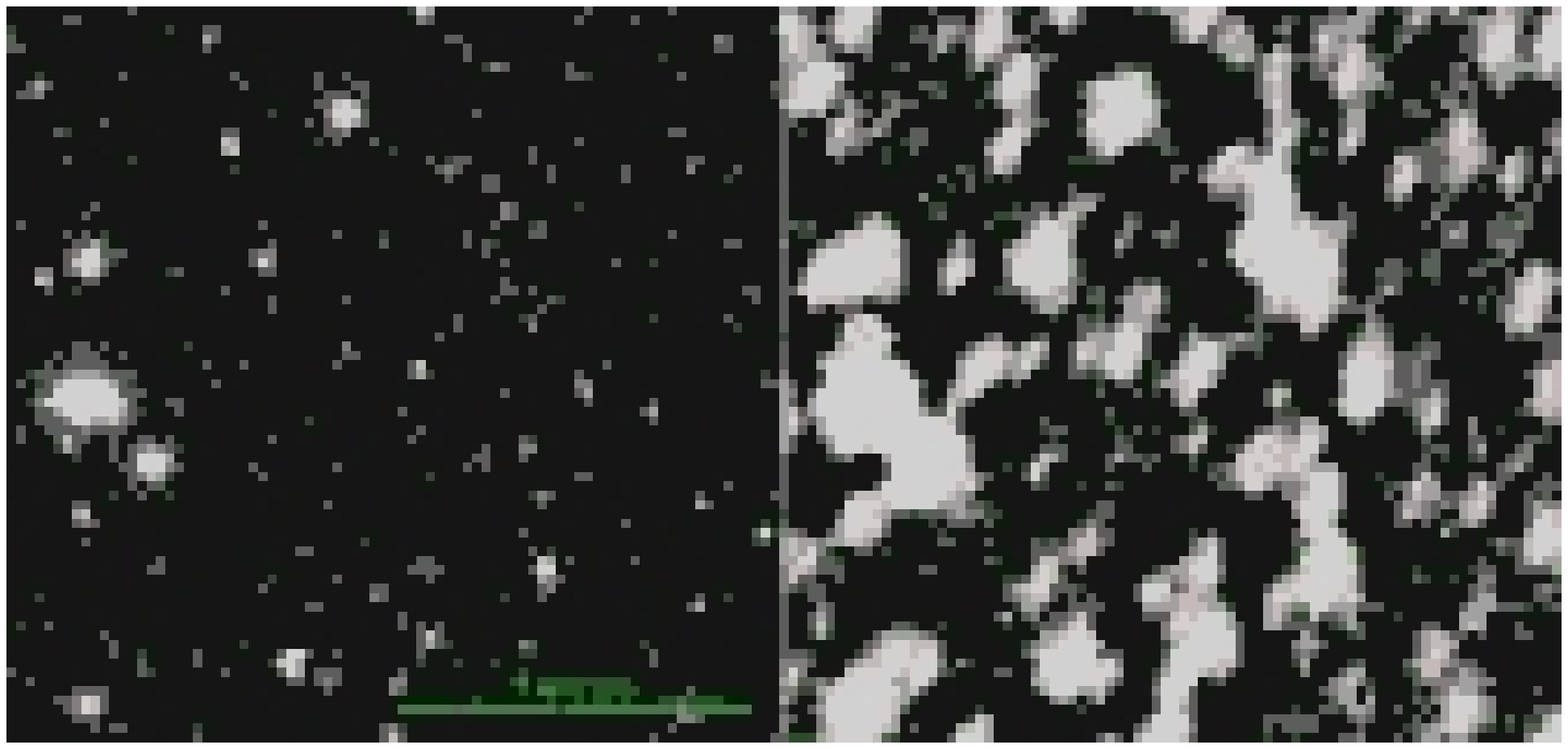}
\FigureFile(75mm,80mm){080822_nep_photoz.ps_pages22}
%
%
%
%
%
%
%
%
%
%
%
%
%
\end{center}
\caption{(Continued). 
}
\label{fig:images2}
\end{figure*}
\setcounter{figure}{0}
\begin{figure*}
\begin{center}
\FigureFile(90mm,80mm){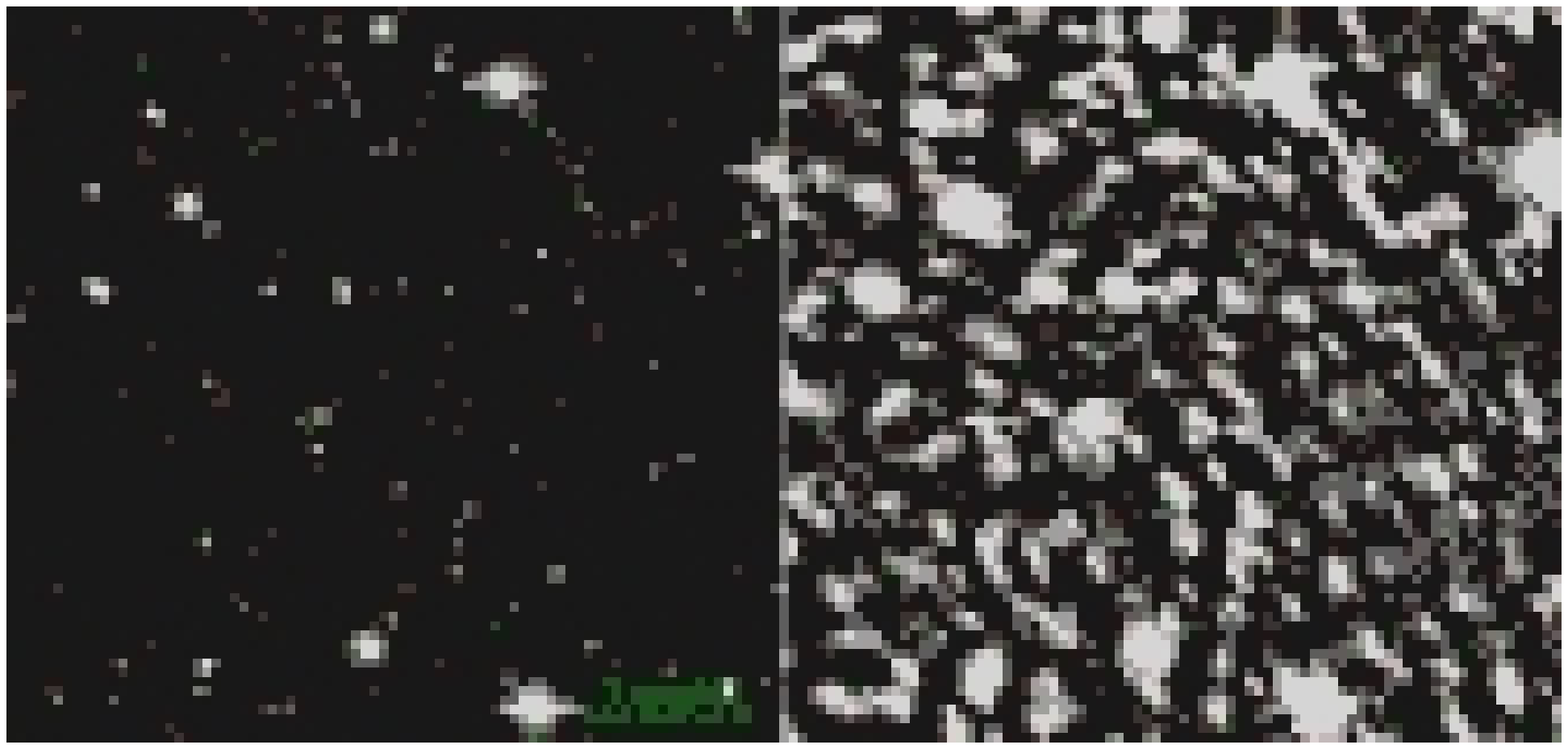}
\FigureFile(75mm,80mm){080822_nep_photoz.ps_pages23}
\FigureFile(90mm,80mm){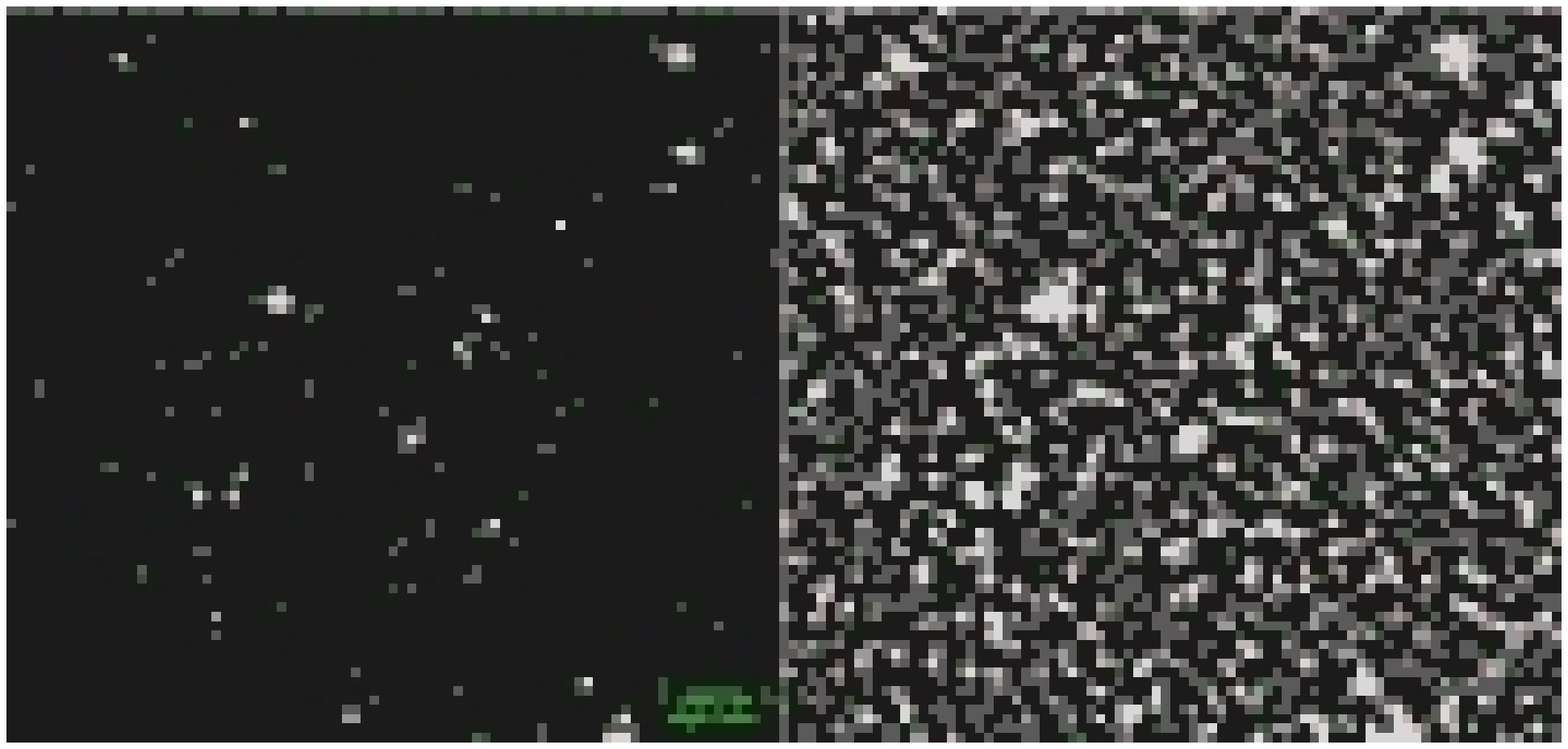}
\FigureFile(75mm,80mm){080822_nep_photoz.ps_pages24}
\FigureFile(90mm,80mm){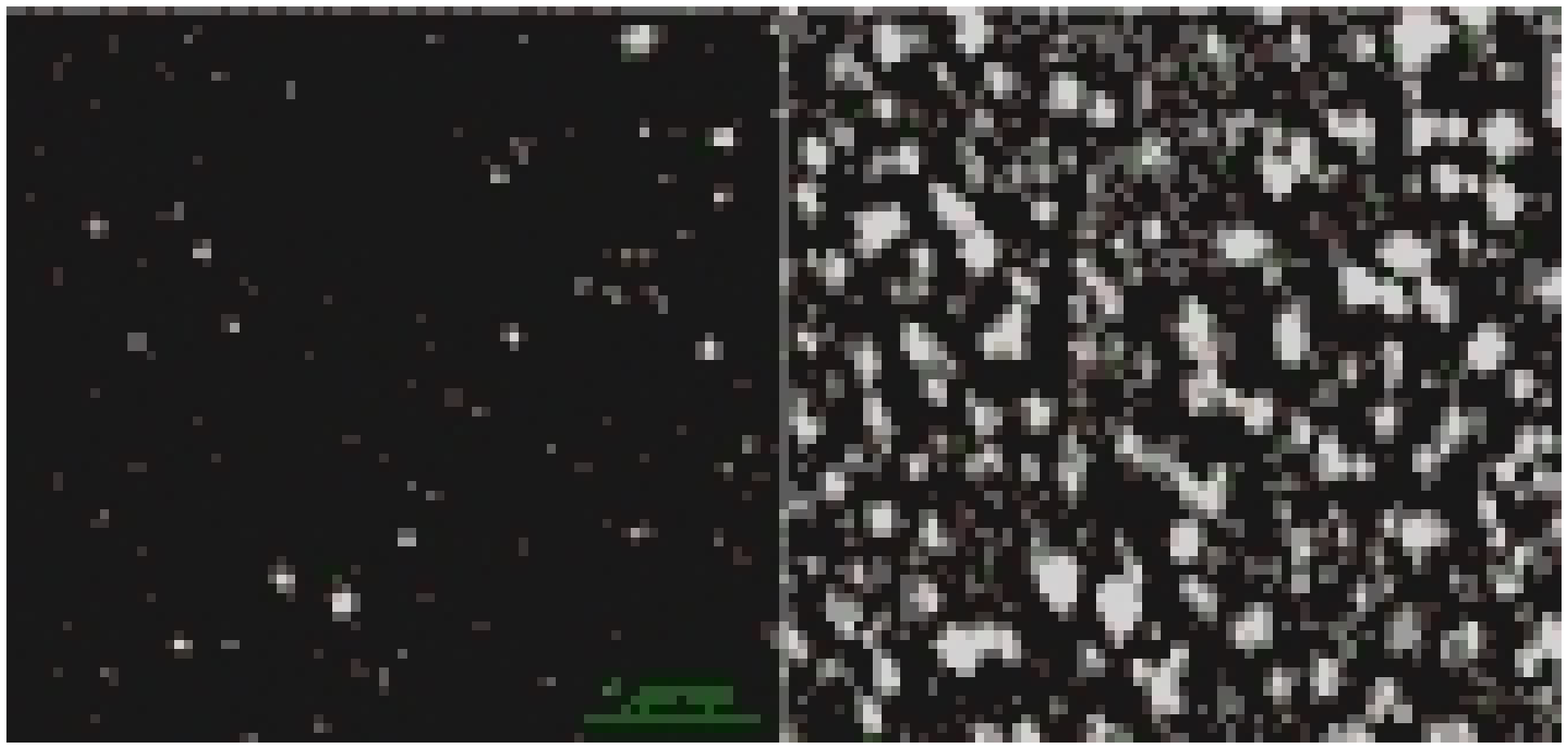}
\FigureFile(75mm,80mm){080822_nep_photoz.ps_pages25}
\FigureFile(90mm,80mm){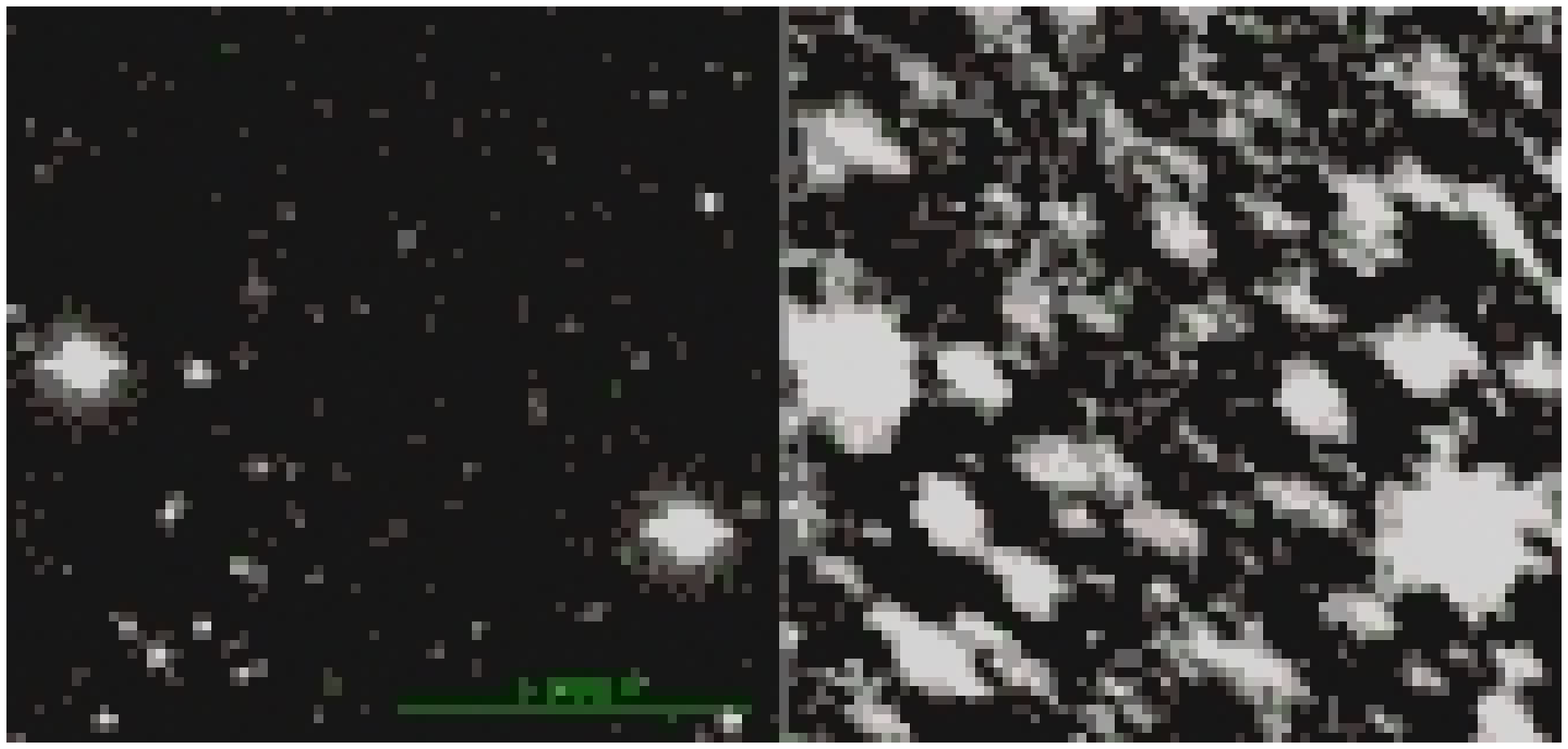}
\FigureFile(75mm,80mm){080822_nep_photoz.ps_pages27}
\end{center}
\caption{(Continued). 
}\label{fig:images2}
\end{figure*}
\setcounter{figure}{0}
\begin{figure*}
\begin{center}
\FigureFile(90mm,80mm){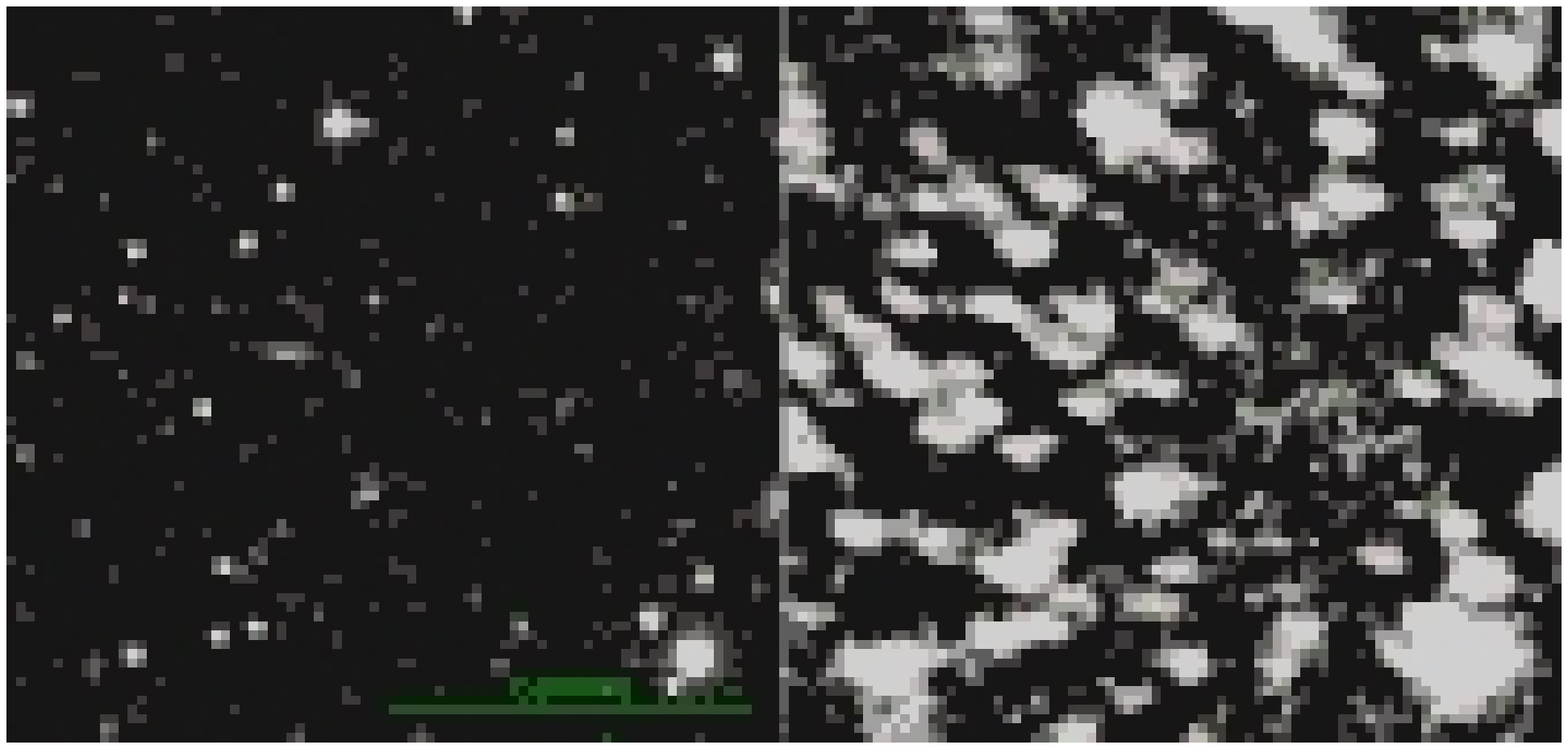}
\FigureFile(75mm,80mm){080822_nep_photoz.ps_pages28}
\FigureFile(90mm,80mm){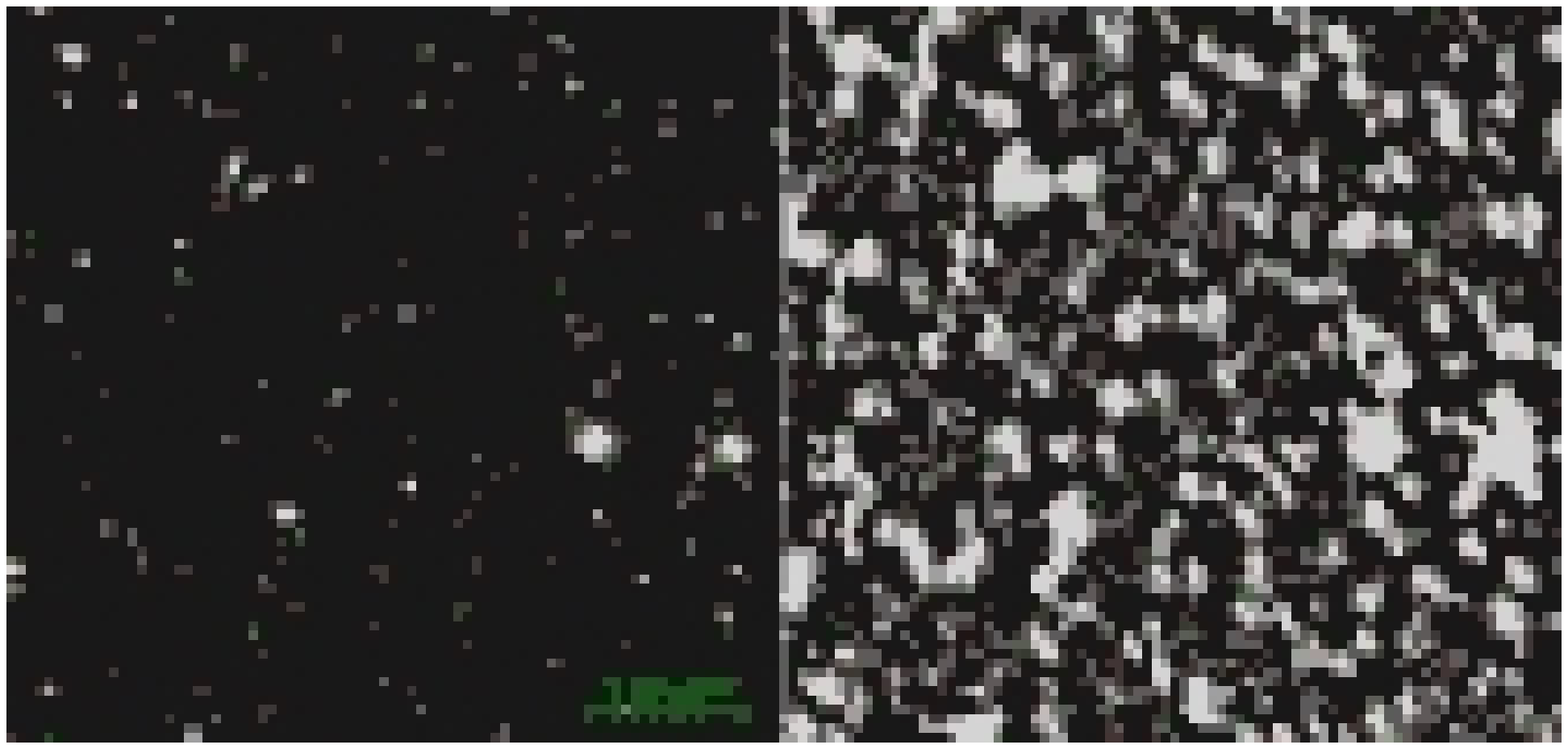}
\FigureFile(75mm,80mm){080822_nep_photoz.ps_pages29}
\FigureFile(90mm,80mm){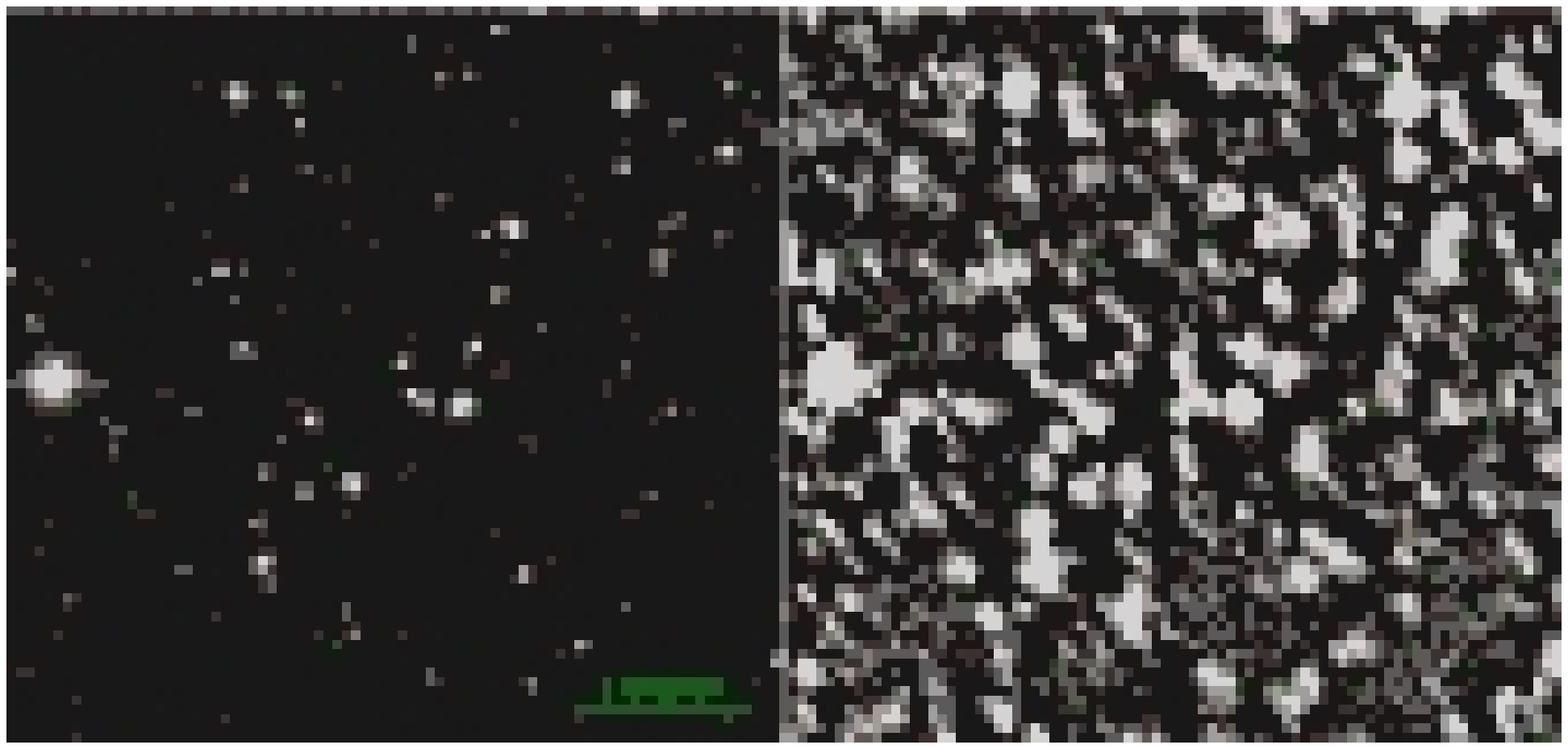}
\FigureFile(75mm,80mm){080822_nep_photoz.ps_pages30}
\FigureFile(90mm,80mm){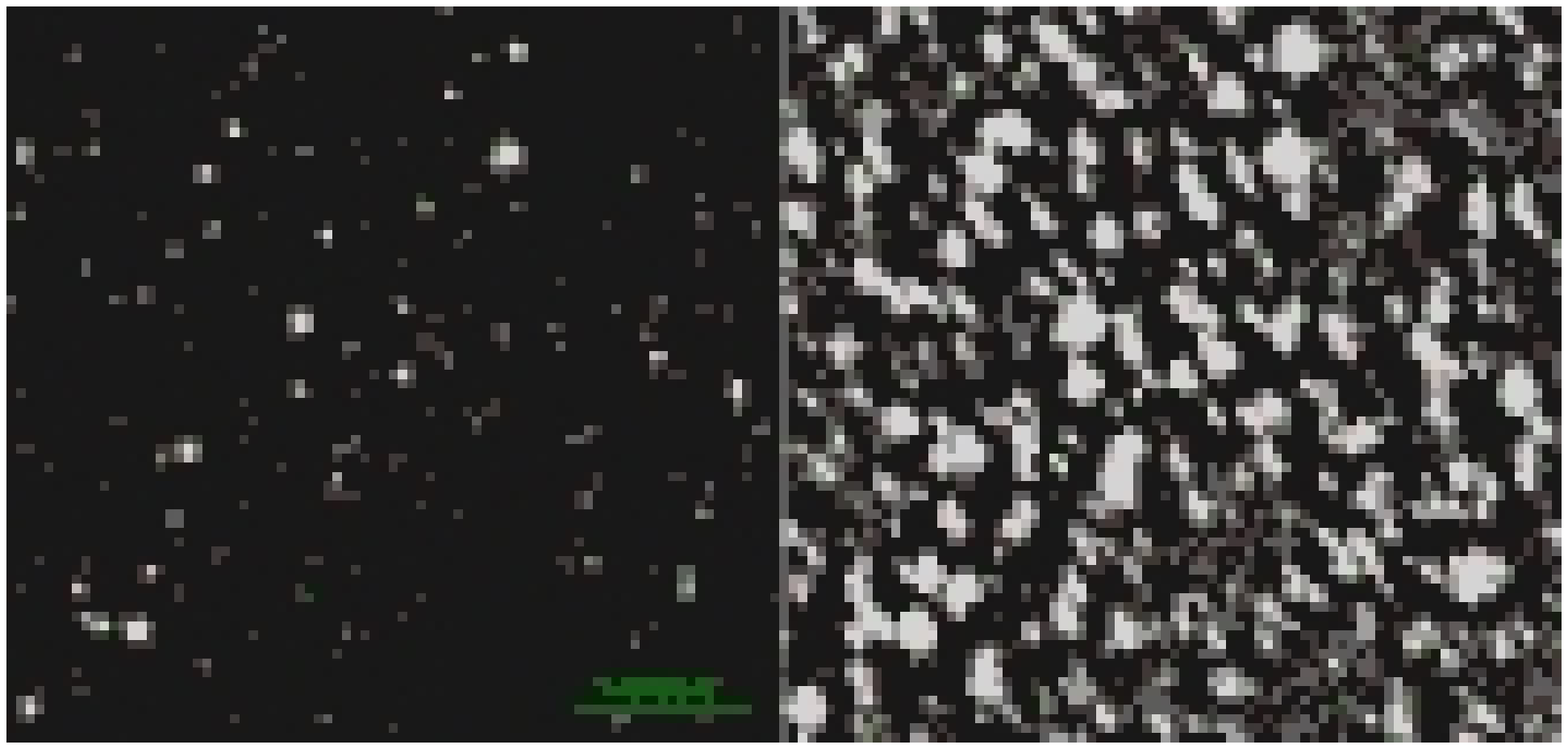}
\FigureFile(75mm,80mm){080822_nep_photoz.ps_pages35}
\end{center}
\caption{(Continued). 
}\label{fig:images2}
\end{figure*}
\setcounter{figure}{0}
\begin{figure*}
\begin{center}
\FigureFile(90mm,80mm){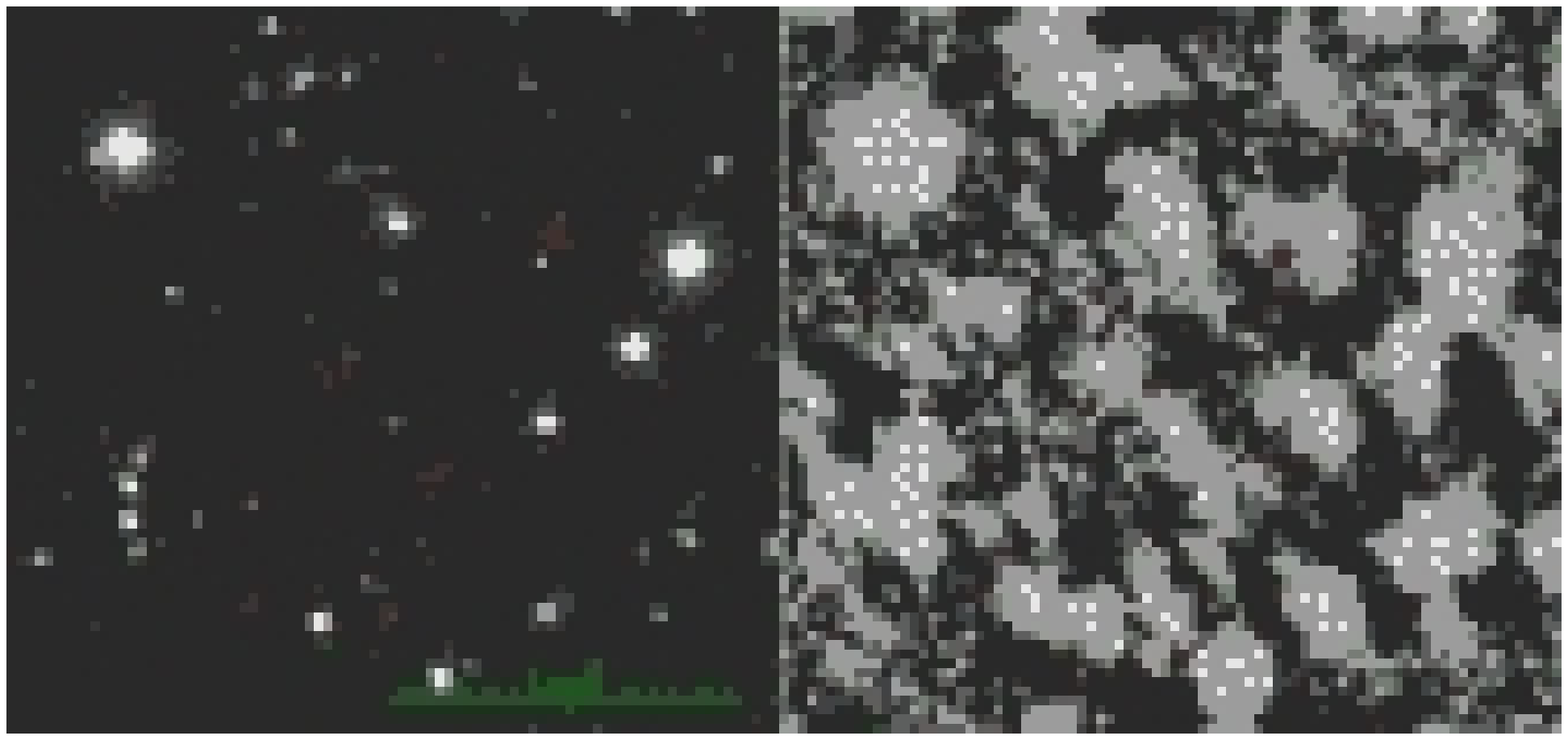}
\FigureFile(75mm,80mm){080822_nep_photoz.ps_pages36}
\FigureFile(90mm,70mm){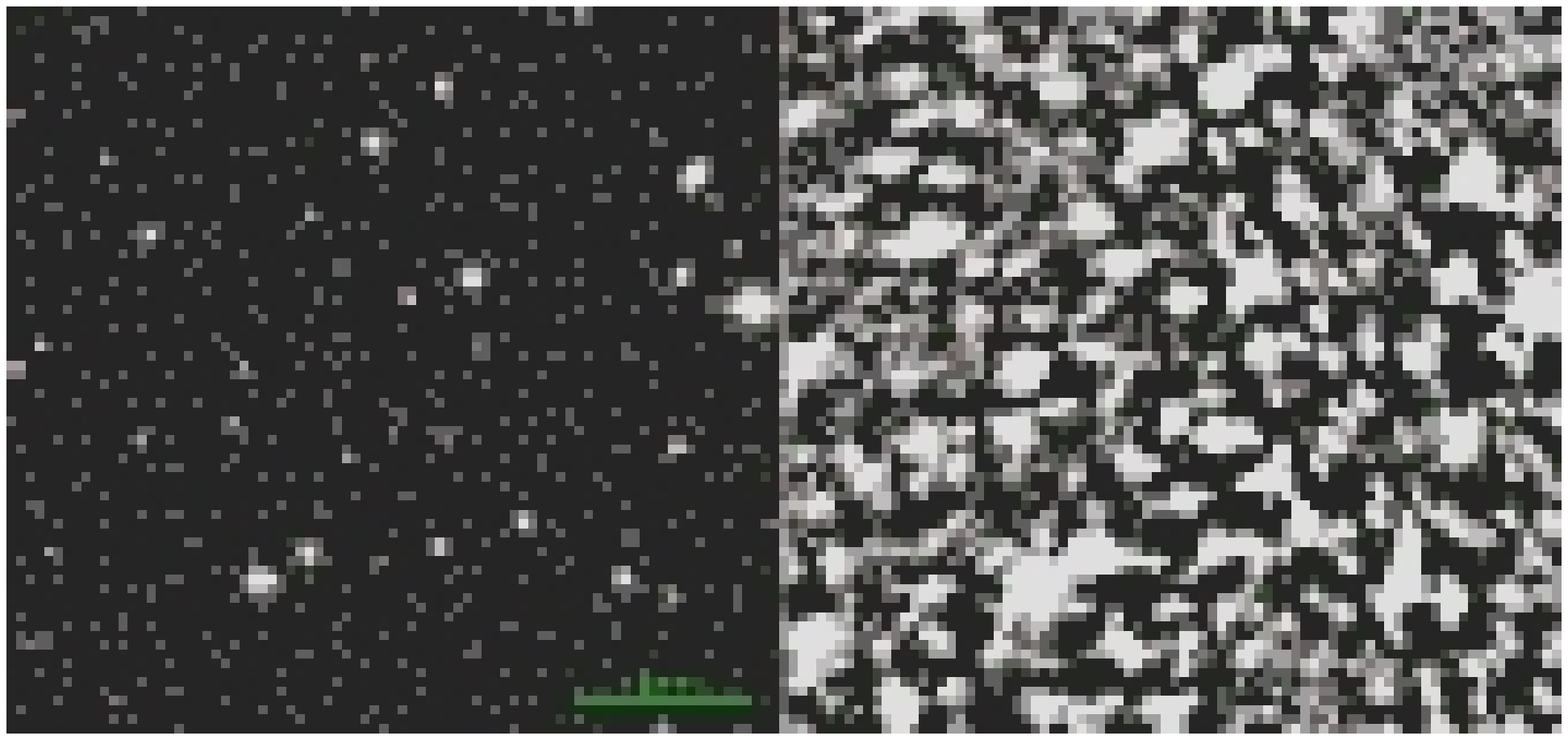}
\FigureFile(75mm,70mm){080822_nep_photoz.ps_pages37}
\end{center}
\caption{(Continued). 
}\label{fig:images2}
\end{figure*}

\begin{figure}
  \begin{center}
    \FigureFile(100mm,100mm){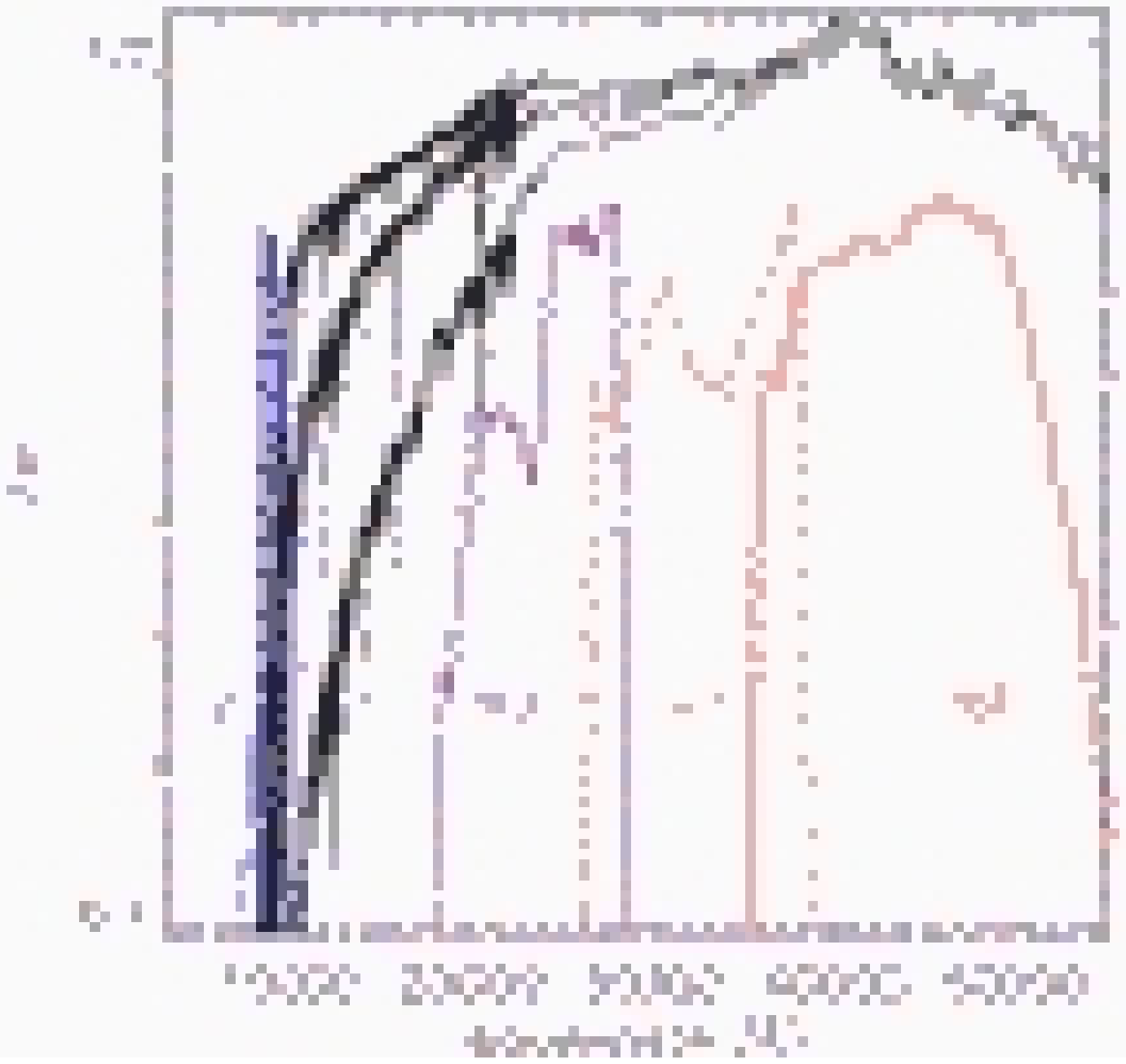}
  \end{center}
  \caption{
Spectra of galaxies with ages of 0.5, 1 and 10 gigayears redshifted to z=1.3. The spectra are from the instantaneous burst model of \citet{2003MNRAS.344.1000B} using the solar metallicity and the Salpeter initial mass function.   Overlaid are the filter response curve of $z',N2,N3,$ and $N4$. 
}\label{fig:filters}
\end{figure}

\begin{figure*}
  \begin{center}
\FigureFile(180mm,180mm){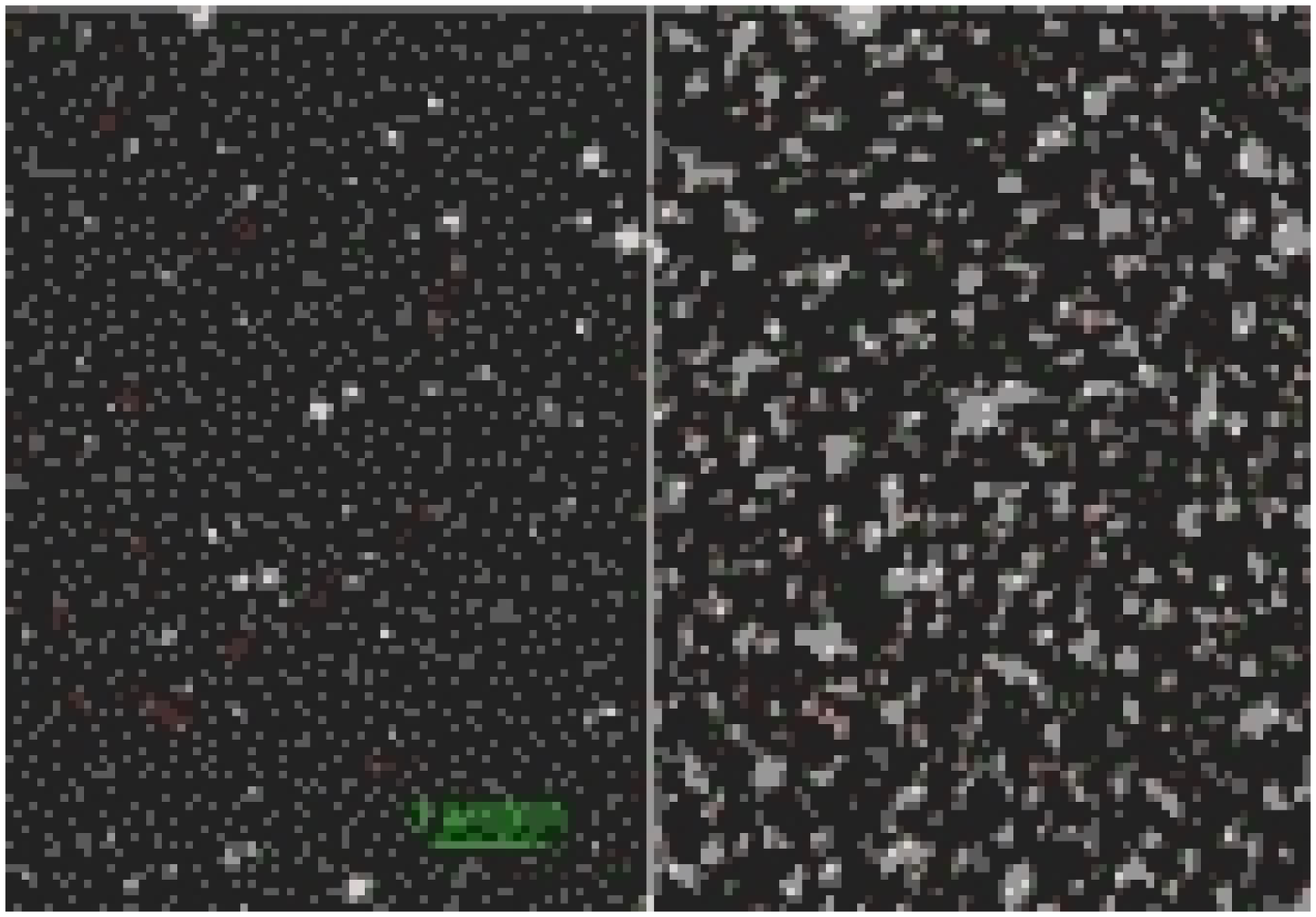}
 \end{center}
  \caption{
The left panel shows the $z'$-band image of a known galaxy cluster RXJ175719.4+663131. 
The scale is indicated in the figure.
The right panel shows the $N4$-band image of the same region. 
Galaxies marked with red circle have consistent $z'-N4$ color to be at z=0.6909.
}\label{fig:RXJ1757_image}
\end{figure*}
\begin{figure}
  \begin{center}
\FigureFile(100mm,100mm){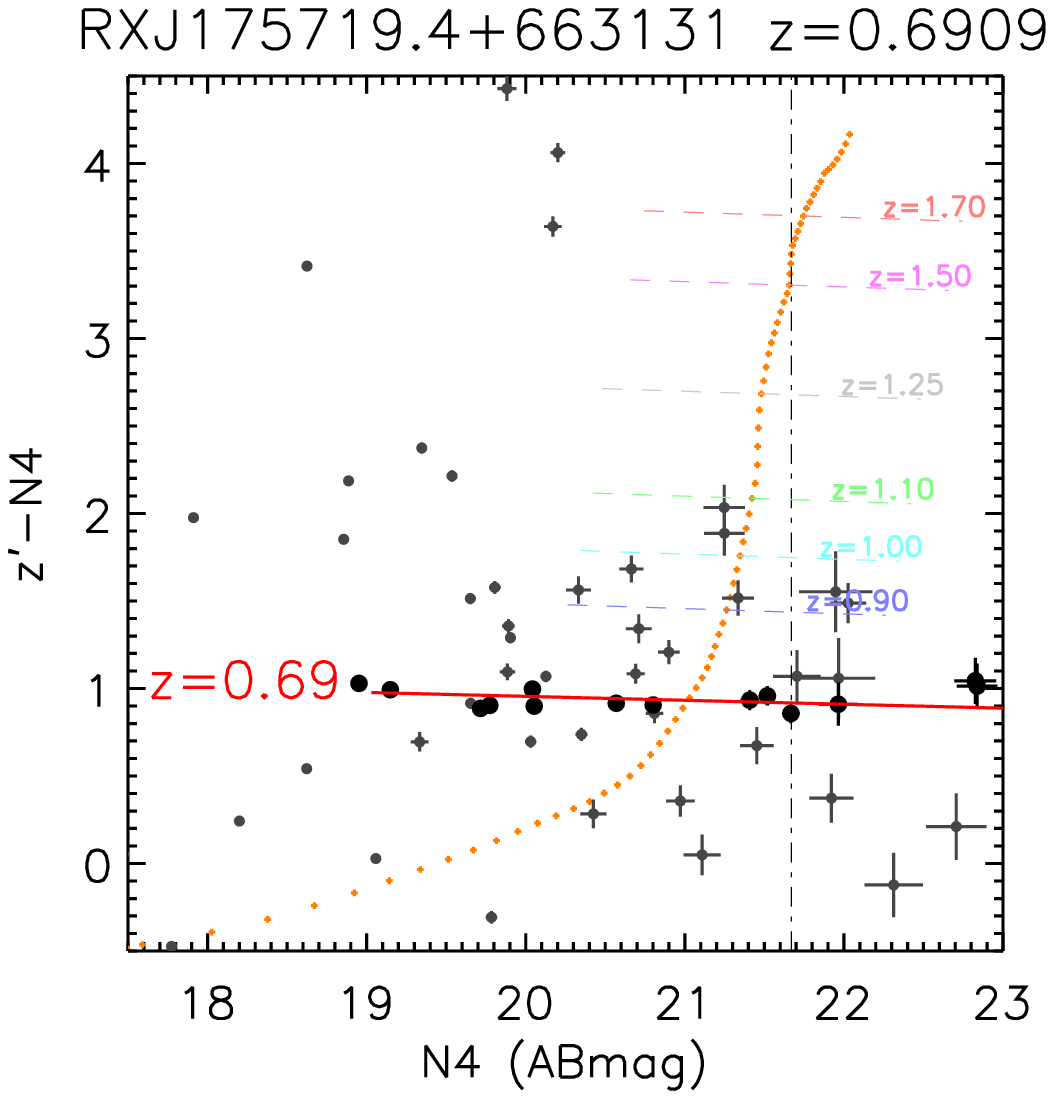}
 \end{center}
  \caption{The $z'$ vs $z'-N4$ color-magnitude relation of the RXJ175719.4+663131. 
 Galaxies plotted with the large dots have consistent $z'-N4$ color to be at z=0.6909 and within 0.5 Mpc from the cluster center.
 The small dots are all galaxies within  within 0.5 Mpc from the cluster center.
 The red line is an expected location of the color-magnitude relation at z=0.6909.
 The orange dots show model galaxy colors with evolution as a function of redshift.
}\label{fig:RXJ1757_cmd}
\end{figure}

\subsection{S15/S9 flux ratio}

 In this subsection, we investigate star-formation activity of cluster galaxies found in Section \ref{sec:15clusters} using the PAH emissions redshifted into $15 \mu$m-band. The 5 $\sigma$ sensitivity in the AKARI $S9W$ and $L15$ bands are 58 and 117 $\mu$Jy \citep{WadaNEP2008}. With these sensitivities, the lower limit of the luminosity range we explore at $0.9<z<1.7$ are approximately $>2-9\times 10^{11}L_{\odot}$ and $>1-7\times 10^{11}L_{\odot}$ in 9 and 15$\mu$m, respectively.


 In this redshift range of $0.9<z<1.7$, star-forming galaxies are known to have a large $S_{15\mu m}/S_{9\mu m}$ flux ratio since prominent PAH emissions at 6.2,7.7, and 8.6$\mu m$ are redshifted into the $L15$-band and form a pseudo-break to the continuum at $<5\mu m$ (see \cite{2007PASJ...59S.557T,takagi_subaru} for a model prediction, and Ohyama et al. 2008 in prep. for spectroscopic confirmation). And of course, the mid-infrared $S_{15\mu m}/S_{9\mu m}$ flux ratio is free from dust extinction unlike optical.
 As a star formation indicator, we prefer the $S_{15\mu m}/S_{9\mu m}$ flux ratio over the mere luminosity in $L15$, since $15\mu$m luminosity largely depends on galaxy mass, and thus, is not an ideal tracer of the star formation activity when targetting massive galaxies in clusters \citep{2002PASJ...54..515G,2005MNRAS.359.1415G}. In the presence of warm dust of the AGN, it is difficult to tell if the $L15$ luminosity is from the AGN or the star-formation.   On the other hand, the $S_{15\mu m}/S_{9\mu m}$ flux ratio does not depend on the mass (luminosity) of galaxies,  analogously to the specific SFR or SFR per unit mass. 
 Therefore, a census of star-formation activity can be performed using the $S_{15\mu m}/S_{9\mu m}$ flux ratio.

\begin{figure}
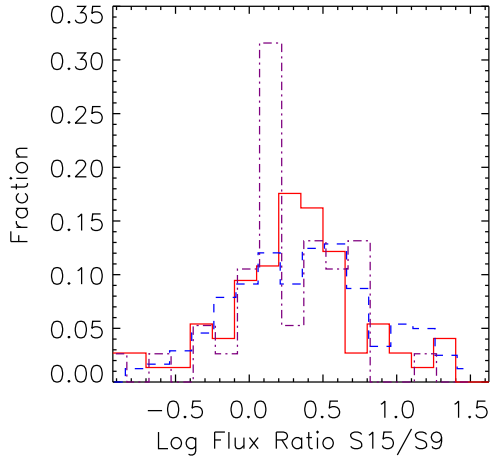

  \begin{center}
\FigureFile(100mm,100mm){080922_nep_photoz.ps_pages}
 \end{center}
  \caption{Normalized distribution of S15/S9 flux ratio. All histograms are normalized so that fractions add up to unity. 
The red solid line is for red-sequence cluster galaxies in the redshift range of  $0.9<z<1.7$.
The blue dashed line is for field galaxies in $0.9<z<1.7$ selected using the photo-z. 
The purple dash-dotted line is for cluster galaxies in $0.9<z<1.7$ selected using the photo-z. 
The median  $S_{15\mu m}/S_{9\mu m}$ ratios are 1.9, 1.4 and 2.3 for the cluster red-sequence, photo-z cluster, and field galaxies, respectively. 
}\label{fig:s15/s9}
\end{figure}

  In Fig.\ref{fig:s15/s9}, we show normalized histograms of $S_{15\mu m}/S_{9\mu m}$ flux ratios. 
 The red solid line is for cluster member galaxies, which were selected in Section \ref{sec:15clusters} as galaxies in $\delta (z'-N4)<0.2$ from the best-fit color of the red-sequence, and within 1.5 Mpc from the cluster center. All histograms are normalized so that fractions add up to unity. 
 The blue dashed line is for field galaxies in the same redshift range ($0.9<z<1.7$) selected using the photometric redshift computed by \citet{takagi_subaru} and \citet{WadaNEP2008}. Briefly, we used the hyperz code with GALEX $FUV,NUV,B,V,R,I,z,J$, and $K$ as inputs whenever available. The resulting photometric redshift estimates agree reasonably well with a dozen of galaxies with spectroscopic redshift in the NEP field \citep{takagi_subaru}.  
 We also show galaxies with photometric redshift consistent to be at a cluster redshift in Fig.\ref{fig:images} with the green triangles. 
  Some photo-z selected galaxies are redder than the red-sequence. These may be  dusty-galaxies or errors in photometric redshift. Identities of these need to be verified spectroscopically. 
 The purple dotted-dash line is photo-z selected cluster galaxies which have consistent photometric redshits with that of the cluster itself and within 1.5 Mpc of radius.

 Compared with field galaxies at $0.9<z<1.7$, the $S_{15\mu m}/S_{9\mu m}$ distribution of cluster red-sequence galaxies are skewed toward a lower $S_{15\mu m}/S_{9\mu m}$ ratio. The median $S_{15\mu m}/S_{9\mu m}$ ratios are 2.3 and 1.9 for the field and cluster red-sequence galaxies, respectively.  
  The photo-z selected cluster members also have lower ratio of $S_{15\mu m}/S_{9\mu m}$ than the field galaxies. 
  The median $S_{15\mu m}/S_{9\mu m}$ ratio is also lower with 1.4 compared with 2.3 for the field galaxies.  
 According to a Kolomogorov-Smirnov test, the photo-z and red-sequence cluster members are different from the photo-z selected field galaxies with 73 and 88\% of significance.
 The difference indicates that both cluster red-sequence and photo-z galaxies have lower star formation activity even at $0.9<z<1.7$, quite similarly to red, passive galaxy clusters in the local Universe. 
 At the same time, the difference in the $S_{15\mu m}/S_{9\mu m}$ ratio suggests that our clusters are not a coincidence of an alignment of galaxies, but a real clustering of galaxies in redshift space.

\section{Discussion}


 Caution must be taken when the number density of our clusters is statistically discussed. Since we sample galaxies to the detection limit of the AKARI $N4$ magnitude, inevitably a narrower luminosity range (by about half a magnitude) is sampled for higher redshift clusters, i.e., our sample of clusters are not a luminosity nor mass selected sample although our cluster detection is machine-based, and can be corrected for selection effects in principle by using a numerical simulation.  Rather we have attempted to the best of our data to detect as low mass clusters as possible toward higher redshift where optical searches are difficult. 

 The total volume surveyed in the 0.4 deg$^2$ NEP field at $0.9<z<1.7$ is 9.1$\times$10$^5$h$^{-3}$Mpc$^3$.
 In the local universe, the number density of massive clusters ($\sim10^{14}M_{\odot}$) is 3$\times$10$^{-5}$h$^{3}$Mpc$^{-3}$  \citep{2007ApJ...657..183R,2008ApJ...679L...1R}.
 At $z>1$, the cluster number density is about an order smaller than this, assuming the Press-Schechter formulation and the $\Lambda$CDM cosmology \citep{1998ApJ...504....1B,2002ARA&A..40..643C},
 becoming close to  3$\times$10$^{-6}$h$^{3}$Mpc$^{-3}$.
 In Table \ref{tab:LTsample}, we show 16 promising cluster candidates. 
 We have searched for a volume that could include a few massive clusters ($\sim10^{14}M_{\odot}$) in the AKARI NEP deep field. 
 For a comparison, \citet{2006MNRAS.373L..26V} also found 13 clusters in the 0.5 deg$^2$ region of the UKIDSS survey.

 We have cross-correlated our cluster candidates with the positions of ROSAT X-ray detections \citep{2006ApJS..162..304H}. 
However, there was no match except for the RXJ175719.4+6633 at $z$=0.6909. The match proves the validity of  our cluster detection
     algorithm. On the other hand, none of our $z > 0.9$ clusters were
     matched with the known X-ray sources, as is expected since the
     flux-limit of the ROSAT is
$\sim 2\times 10^{-14}$ ergs cm$^{-1}$s$^{-1}$ in the 0.5-2.0 keV band, which requres large X-ray luminosity of $>2 \times 10^{44}$ erg s$^{-1}$ at $z>1$ to be detected with the ROSAT. The expected X-ray luminosity of our clusters are much smaller than this at around $3 \times 10^{43}$ erg s$^{-1}$ assuming the mass of $<10^{14}M_{\odot}$ \citep{2007ApJ...668..772M}.

In the literature, there have been a few attempts to locate galaxy clusters at $z>1$ using the optical-near infrared data.
\citet{2007ApJ...664L..17M} found a compact group of massive red galaxies  at $z=1.51$ in 120 arcmin$^2$ of the Gemini Deep Deep survey.
The UKIDSS Ultra Deep Survey detected 13 cluster candidates with $0.61 \leq z \leq 1.39$ in a 0.5 deg$^2$ survey \citep{2005ApJ...634..861Y,2006MNRAS.373L..26V}. 
\citet{2007ApJ...671.1497C} found an overdensity of galaxies at z=1.6, presumably a poor cluster in the process of formation, in the GOODS-MUSIC catalog using photometric redshifts only.  
However, these surveys still suffer from relatively smaller field of view of the near-infrared detectors of ground-based telescopes. A major step forward will perhaps be brought by infrared space telescopes such as the Spitzer space telescope and our AKARI satellite, which can survey large area faster without suffering from the atmospheric background in the infrared.
\citet{2008arXiv0804.4798E} have identified 106 galaxy cluster candidates at $z>1$ using a 4.5 $\mu$m selected sample of objects from a 7.25 deg$^2$ region in the Spitzer Infrared Array Camera (IRAC) Shallow Survey and the NOAO Deep Wide Survey area.  They have spectroscopic confirmation for 12 clusters out of the 106.
 Although our AKARI NEP DEEP survey is 0.4 deg$^2$, cluster finding in the AKARI NEP wide survey  in 6 deg$^2$ \citep{2006PASJ...58..673M,2007PASJ...59S.543M,2007PASJ...59S.529L} is underway with somewhat shallower detection limit. In addition, it is our advantage that the AKARI fields have photometry in 9 consecutive passbands in 2-24$\mu m$ ($N2,N3,N4,S7,S9W,S11,L15,L18W$ and $L24$), allowing us to immediately investigate near-, mid-infrared properties of cluster galaxies.

%


Our detection of no enhanced PAH emission in terms of $S_{15\mu m}/S_{9\mu m}$ emission in  $0.9<z<1.7$ clusters is quite interesting. \citet{2006MNRAS.370..198C} found that among blue galaxies, more luminous galaxies are in the dense regions at $z\sim 1$. 
 \citet{2007A&A...468...33E} showed that the star formation-density relation was reversed at z$\sim$1: the average SFR of an individual galaxy increased with local galaxy density at z$\sim$1. 
Recently, Koyama et al. (2008) found  possible dusty-starforming galaxies detected in 15$\mu$m in a galaxy cluster at $z\sim 0.8$.   
\citet{2008MNRAS.386.1907S} showed that at least a third of the Spitzer-identified submm galaxies at $1<z<1.5$ appear to reside in over-densities, supporting the high-redshift reversal of the local starformation-density relation. 

These results indicate that galaxies in the dense regions are rapidly evolving at z$\sim$1. In the local Universe, the galaxy SFR is lower at dense regions \citep{2003ApJ...584..210G,2004AJ....128.2677T}.
According to the downsizing  galaxy formation scenario, where stellar growth systematically moves to lower mass systems \citep{1996AJ....112..839C},  more massive galaxies are expected to be star formating at higher redshift\citep{2004MNRAS.350.1005K,2005ApJ...622L...5T,2006A&A...453L..29C,2006MNRAS.372..933N}. In this case, one might expect galaxies in dense regions have stronger PAH emission at $z\sim$1. Our finding, however, is opposite; cluster red-sequence and the photo-z selected cluster galaxies are already more passive even at $0.9<z<1.7$. The formation epoch of these passive red-sequence galaxies must be at least a few Gyrs further in the past from $0.9<z<1.7$. At the same time, our results suggest that galaxy formation epoch is quite earlier in the dense regions than in the field region, i.e., our results require even stronger downsizing galaxy formation \citep{1996AJ....112..839C}, and possibly implying the presence of separate physical mechanisms responsible in the dense/sparsely populated  regions \citep{2003MNRAS.346..601G}.




%
%


\section{Summary}

 The deep $N4$ image with the AKARI, the first Japanese satellite dedicated to infrared astronomy, combined with the $z'$-band imaging with the Subaru telescope provides us with an opportunity to find massive galaxy clusters at $0.9<z<1.7$, where previous searches for galaxy clusters based on optical photometry were difficult. By applying an efficient galaxy cluster finder we developed, we have selected 16 promising galaxy cluster candidates at $0.9<z<1.7$ in the AKARI NEP deep field.

 Our cluster candidates show an obvious over-density of galaxies in the Subaru $z'$-band image, in addition to a tight color-magnitude relation in $z'-N4$ vs. $N4$ (Fig.\ref{fig:images}). And thus, they are very likely to be a true clustering of galaxies at $0.9<z<1.7$.

 Using these high redshift galaxy clusters, we have investigated the distribution of the $S_{15\mu m}/S_{9\mu m}$ flux ratio, which is an extinction-free indicator of star-formation activity at  $0.9<z<1.7$.
 Compared with photo-z selected field galaxies in the same redshift range, both the red-sequence and photo-z selected cluster member galaxies statistically show a lower $S_{15\mu m}/S_{9\mu m}$ flux ratio. This result indicates that cluster member galaxies have less star-formation activity even at $0.9<z<1.7$, suggesting that the formation epoch of these cluster galaxies are at higher redshift.

%
%
%
%


\begin{longtable}{rrrrrr}
  \caption{List of galaxy cluster candidates detected in the NEP deep field. 
  $\delta_{\Sigma}$ is a local overdensity.  $\sigma_{\Sigma}$ is a significance of the overdensity (see text for details.)
}\label{tab:LTsample}
  \hline              
  Name (coordinates) & z$_{photo}$ & $N_{member}$ & $\delta_{\Sigma}$  & $\sigma_{\Sigma}$\\
\endfirsthead
  \hline
  name & value & value2 & 1 \\
\endhead
  \hline
\endfoot
  \hline
\endlastfoot
  \hline
J175348.9+662422 & 0.94 &      14 &  4.0 & 2.2\\
J175712.5+663114 & 0.92 &      21 &  4.0 & 2.4\\
J175341.9+664341 & 0.96 &      10 &  3.4 & 1.6\\
J175639.6+664007 & 1.07 &      12 &  4.0 & 1.6\\
J175513.4+664254 & 1.03 &      16 &  3.1 & 1.2\\
J175708.2+664351 & 1.07 &      19 &  8.9 & 6.3\\
J175327.2+662351 & 1.13 &      14 &  7.5 & 4.6\\
J175630.2+663849 & 1.14 &      19 &  5.2 & 2.8\\
J175448.8+664900 & 1.17 &      12 &  4.6 & 2.2\\
J175323.4+665045 & 1.19 &       8 &   5.7 & 3.2\\
J175445.2+662640 & 1.23 &      14 &   5.6 & 2.4\\
J175648.3+662216 & 1.24 &       7 &    6.8 & 3.6\\
J175621.3+663357 & 1.29 &      14 &    7.9 & 3.6\\
J175558.5+663433 & 1.36 &      13 &    6.8 & 3.1\\
J175612.5+664005 & 1.50 &       5 &    10.0 & 3.9\\
J175653.6+663303 & 1.67 &       7 &     6.6 & 3.4\\
\end{longtable}





\section*{Acknowledgments}
We thank the anonymous referee for many insightful comments, which significantly improved the paper.
T.G. acknowledges financial
support from the Japan Society for the Promotion of
Science (JSPS) through JSPS Research Fellowships for Young
Scientists.
MI was supported by Creative Research Initiatives (CEOU)
 of MEST/KOSEF


\end{document}